\documentclass[11pt]{article}
\pdfoutput=1
\usepackage[utf8]{inputenc}

\usepackage{amstext,amsmath,amssymb,amsfonts,bbm}
\usepackage{hyperref}
\usepackage{authblk}
\usepackage{epsfig} 
\usepackage{xcolor} 
\usepackage{graphicx} 
\usepackage{braket} 
\usepackage{slashed} 
\usepackage{dsfont} 
\usepackage{amsthm}  
\usepackage{physics}
\usepackage{cancel}
\usepackage{cleveref}

\usepackage[normalem]{ulem} 

\usepackage{adjustbox}
\usepackage{tabularx}
\usepackage[skip=1ex]{caption}
\usepackage{subcaption}
\usepackage{textcomp}

\usepackage{cite}

\allowdisplaybreaks[4]

\usepackage[lmargin=50pt,rmargin=60pt,tmargin=60pt,bmargin=65pt]{geometry}

\newcommand{\be}{\begin{equation}}
\newcommand{\ee}{\end{equation}}

\newcommand{\psib}{\bar{\psi}}

\allowdisplaybreaks[4]

\theoremstyle{remark}

\begin{document}

\begin{titlepage}

\title{
\hfill\parbox{3cm}{ \normalsize YITP-25-156}\\   
\vspace{1cm} 
Characteristic polynomials of 
tensors 
via Grassmann integrals
and distributions of roots for random Gaussian 
tensors
}

\author[1]{Nicolas Delporte}
\author[1]{Giacomo La Scala}
\author[2, 3]{Naoki Sasakura}
\author[1]{Reiko Toriumi}

\affil[1]{\normalsize\it Okinawa Institute of Science and Technology Graduate University, 1919-1, Tancha, Onna, Kunigami District, Okinawa 904-0495, Japan. \hfill}
\affil[2]{\normalsize\it 
Yukawa Institute for Theoretical Physics, Kyoto University, Kitashirakawa, Sakyo-ku, Kyoto 606-8502, Japan.\hfill}
\affil[3]{\normalsize\it 
CGPQI, Yukawa Institute for Theoretical Physics, Kyoto University,
Kitashirakawa, Sakyo-ku, Kyoto 606-8502, Japan.
\authorcr
Emails: {\rm\url{nicolas.delporte@oist.jp}, \url{giacomo.lascala@oist.jp}, \url{sasakura@yukawa.kyoto-u.ac.jp}, \url{reiko.toriumi@oist.jp}.} \authorcr \hfill }

\date{}

\maketitle

\begin{abstract}
\noindent 
We propose a new definition of characteristic polynomials of tensors based on a partition function of Grassmann variables.
This new notion of characteristic polynomial addresses general tensors including totally antisymmetric ones, but not totally symmetric ones.
Drawing an analogy with matrix eigenvalues obtained from the roots of their characteristic polynomials, we study the roots of our tensor characteristic polynomial.
Unlike standard definitions of eigenvalues of tensors of dimension $N$ giving 
$\sim e^{{\text{constant}} \,  N}$ 
number of eigenvalues, our polynomial always has $N$ roots.
For random Gaussian tensors, the density of roots follows a generalized Wigner semi-circle law based on the Fuss-Catalan distribution, introduced previously by Gurau [arXiv:2004.02660 [math-ph]].
\end{abstract}

\end{titlepage}

\tableofcontents

\section{Introduction}

Tensors enjoy different definitions of eigenvalues (the most common ones being E, Z and H) \cite{qi2005eigenvalues, qi2018tensor} playing roles in different contexts  generalizing matrix spectral theory to higher-order operators, with applications in data analysis \cite{pollock2023accelerating}, optimization theory \cite{hu2013finding}, hypergraph theory \cite{galuppi2023spectral}, quantum entanglement \cite{weinbrenner2025quantifying}, etc. It is still an area of active research to understand the information (algebraic, combinatorial or geometric) about the tensor contained in those eigenvalues and how to appropriately decompose a tensor to fit best one's purpose, e.g. \cite{holweck2022toward}. As the roots of the characteristic polynomials of matrices give their eigenvalues, analogous tensor characteristic polynomials have been introduced \cite{li2012characteristic, hu2013determinants, galuppi2023characteristic} whose roots are associated with the different notions of eigenvalues. So far, although some works consider Z- and H-eigenvalues of generic non-symmetric tensors, e.g. \cite{nie2018real,benson2019computing}, attention has mainly been focused on symmetric tensors. 

Additionally, including randomness is key to simplifying equations or to lead to explicit solutions. The study of characteristic polynomials of random matrices has brought indications of universality (see e.g. \cite{akemann2021characteristic}) as well as rich connections with other fields, for example, deep links with number theory indicate relations between the distribution of zeros of characteristic polynomials of random matrices and the zeros of the Riemann zeta function \cite{keating2000random,brezin2000characteristic}. 
To our knowledge, an analogous program for random tensors is still in its infancy. We intend here to broaden the set of tools that are typically used to decompose tensors, emphasizing some universal aspects that emerge from tensors.

In this work, we will introduce a new notion of tensor characteristic polynomial based on a Grassmann integral (for physicists, a partition function of a zero-dimensional fermionic system), suitable for totally antisymmetric tensors, but in principle fitting all permutation symmetries of the tensor indices except for the fully symmetric one. To be able to write explicit equations, we will consider Gaussian tensors. We will show that under that condition and in the limit of large dimension, the Grassmann integral can be evaluated with a saddle-point method and the distribution of its roots shown to follow a Fuss-Catalan distribution. In this way, we bring a new point of view to the close relation between random tensors and Fuss-Catalan distributions, in parallel to the previous work \cite{gurau2020generalizationwignersemicirclelaw} at the heart of the concept of freeness for tensors \cite{bonnin2024freeness,bonnin2024tensorial,bonnin2024universality,nechita2025tensorfreeprobabilitytheory,collins2024free}. 

In the following, we will recall in Sec.~\ref{sec:matrices} the use of Grassmann integrals for determining the spectrum of matrices through their characteristic polynomial. In Sec.~\ref{sec:tensors}, we will present our definition of tensor characteristic polynomial and its relation to the hyperpfaffian. In Sec.~\ref{sec:distributionroots}, after averaging the characteristic polynomial over Gaussian tensors, we will evaluate the distribution of its roots at large $N$. Finally, the section~\ref{sec:conclusion} ends with a few remarks pointing to future directions.

\section{Matrices and Grassmann integrals}
\label{sec:matrices}

Determining the spectrum of a matrix brings essential information about its 
properties.
It characterizes invariants under similarity transformations (the determinant and the trace), gives a decomposition of the matrix important for matrix multiplication, specifies the stability of a system modelled with that matrix, etc. One way to determine the spectrum of a matrix is to solve for its eigenvalues and eigenvectors together. Another approach, that does not refer to a particular basis of the vector space that the matrix acts upon, solves for the roots of the characteristic polynomial of the matrix. 

When the matrix is sampled from an ensemble, it is equally important to determine properties of the spectrum and in certain cases, its limiting distribution for matrices of large dimension can be calculated. Edwards and Jones \cite{Edwards_1976} rephrased the problem of calculating the spectral density of random matrices as the evaluation of the free energy of a disordered system, the matrix entering as the coupling between the different degrees of freedom at each site \footnote{
One assumes to give a small imaginary part to $\lambda$, see App.~\ref{app:EJformula}.
}
\be
\label{eq:rhoMat}
\rho(\lambda)=\frac{1}{\pi N}\Im \dv{}{\lambda}\expval{\log \det(\lambda 
\mathds{1} 
-M)}_M\,,
\ee
relying on insights from the statistical physics of glasses, such as the replica and cavity methods. Additionally, many situations (such as for large dimensions $N$, at high temperature, in the absence of replica-symmetry breaking or on the Nishimori line) allow to pass the expectation through the logarithm, going from a quenched to an annealed average over $M$. Then, the system is said ``self-averaging" or that typical configurations concentrate around the average ones. In that approximation, the computation simplifies drastically.
\cite{Edwards_1976} showed that for symmetric random matrices of dimension $N$ with Gaussian entries of zero mean, the spectral density follows the Wigner semi-circle law at large $N$ and with a non-zero mean, an outlier emerges from the semi-circle.
Kamenev and Mezard \cite{kamenev1999wigner} used instead a Grassmann formulation to recover that the spectral density of Gaussian unitary matrices obeyed the semi-circle law asymptotically at large $N$, without the need of replica symmetry breaking, but they showed that higher order correlations of the spectral density were sensitive to higher levels of replica symmetry breaking.

In order to motivate our new definition of 
tensor characteristic polynomials,
we will first review some of the existing ways of defining eigenvalues for matrices.

\subsection{
Matrix  
characteristic polynomial as a Grassmann integral
}
\label{sec:matrixeigenvaluedet}
The first step is to express 
the determinant appearing in the spectral density 
\eqref{eq:rhoMat}
as arising from an integral over Grassmann numbers.  
For a 
complex
matrix $(M_{ab})_{1\leq a,b\leq N}$,
a complex parameter $\lambda$,
and 
Grassmann variables 
of two species \footnote{
Here, $\bar{\psi}$ does not mean the complex conjugate of ${\psi}$, it is just a notation for a different species.
}
$\{\psi_{a},\psib_a\}_{1\leq a\leq N}$,
the characteristic polynomial (in $\lambda$) of $M$ is given by
\be
\det(\lambda \mathds{1}-M)=
\int \mathcal{D}( \psi,\psib) 
\exp(\sum_{a, b=1}^N \psib_a (\lambda 
\delta_{ab}
-M_{a b})\psi_b )\,,
\label{eq:matrixdet}
\ee
where we assume the standard anti-commutation relations 
\be
\{\psi_a,\psi_b\}=\{\psib_a,\psi_b\}=\{\psib_a,\psib_b\}=0
\,,
\label{eq:grassmannanticommute}
\ee
which imply
\begin{equation}
\psi_a \psi_a = 0\,,
\qquad 
\psib_a \psib_a = 0
\,,
\label{eq:anticommuteimplication}
\end{equation}
and the standard normalization (see App. A of \cite{Caracciolo_2013})
\begin{gather}
\mathcal{D}( \psi,\psib) =\prod_{a=1}^N\dd\psi_a\dd \psib_a\,,\quad
\int \mathcal{D}( \psi,\psib)\;\prod_{a=1}^N\psib_a\psi_a=1\,,
\label{eq:normalization}
\end{gather}
where the order of the appearance of $a$ is respected in the product.
The eigenvalues of the matrix $M$, their spacing and correlation can be determined from its characteristic polynomial.
Namely, we look for the zeros of the polynomial \eqref{eq:matrixdet} to find the spectrum of a matrix $M$. 

One can also directly consider the average of the characteristic polynomial over an ensemble of random matrices. 
For Gaussian real symmetric matrices,
it is known that \footnote{See Prop. 11 in \cite{forrester2006counting} that was taking a purely combinatorial approach with enumeration of matchings, involutions, and associated multinomial coefficients.}
\be 
\label{eq:avDetHermite}
\expval{
\det(\lambda \mathds{1}-M)}_{M}
=
\sigma^N 
{\rm He}_N(\lambda/\sigma)
\,,
\ee
where $N \times N$ matrix $M$ is Gaussian distributed
with a variance $\sigma^2>0$ for its off-diagonal elements 
and where
${{\rm He}}_N$
are the (probabilistic) Hermite polynomials. \footnote{Similar formulas hold for other ensembles like the complex Ginibre ensemble, see eq. 6.9 in \cite{AKEMANN2003532}, also relating their average characteristic polynomial to Hermite polynomials, but with other parameters relative to the asymmetry of the distribution entering in the analog of \eqref{eq:avDetHermite}.
} 
Besides, the generating function 
$\mathcal{X}_N$
of 
the sum of 
the $k$-th powers of the (normalized) zeroes of the Hermite polynomials converges weakly (i.e. in moments) to the generating function of the Catalan numbers 
$C_k$
\cite{kornyik2016wigner}. More precisely, 
\begin{gather}
    \frac{1}{N}\mathcal{X}_N(z/\sqrt{N})\to \sum_{k\geq 0} C_k z^{2k} \quad (0\leq z\leq 1/3)\,,
    \quad C_k=\frac{1}{k+1}\binom{2k}{k}
    \\
    \mathcal{X}_N(z)=\sum_{k\geq 0} \Xi_N(k)z^k
    \,,
    \quad \Xi_N(k)=\sum_{j=1}^N \Big(\xi_j^{(N)}\Big)^k
    \,,
\end{gather}
where $\{\xi^{(N)}_j\}_{1\leq j\leq N}$ are the $N$ roots of the $N$-th Hermite polynomial ${{\rm He}}_N$, 
and
they are related to the $N$ eigenvalues of $M$, $\{\lambda_j^{(N)}\}_{1 \le j \le N}$, by the rescaling $\xi^{(N)}_j =2\sqrt{N}\lambda_j^{(N)}$.

It is also worthwhile to recall here that, given an integer $s\geq 1$, the eigenvalues of the matrix $Y_s^*Y_s$, with $Y_s = X_1\cdots X_s$ and $\{X_i\}_{1\leq i\leq s}$ being complex Ginibre matrices, follow Fuss-Catalan distributions (see App.~\ref{app:FC}) in the limit of large dimension \cite{penson2011product}. They form a determinantal point
process with a correlation kernel that can be expressed in terms of Meijer G-functions that can be interpreted as a multiple orthogonal polynomial ensemble \cite{kuijlaars2014singular} and \cite{neuschel2014plancherel} obtained asymptotic formulas for their characteristic polynomials.

\subsection{Pfaffians of antisymmetric matrices}
In order to compute the determinant of 
an antisymmetric
matrix, one can actually use a half of the number of Grassmann variables as used in \eqref{eq:matrixdet}.
Indeed, 
for an antisymmetric matrix $M$ of even dimension $N$,
one may consider its Pfaffian
\begin{equation}
    \text{pf}(M) :=\int D\psi \exp(-\frac{1}{2}\sum_{a,b=1}^N \psi_a M_{ab} \psi_b)
    =
    \sqrt{\det M}\,,
    \quad
    D\psi =  \prod_{a=1}^N \dd \psi_a
    \,.
    \label{eq:matrixpfa}
\end{equation}
In addition to their 
use
in geometry and topology due to their intrinsic connection 
with 
characteristic classes and the topology of vector bundles \cite{pestun2016review}, Grassmann integral techniques offer powerful tools for 
certain 
combinatorial enumeration, such as expressing partition functions of spin systems and counting spanning forests in graphs and hypergraphs \cite{Sportiello:2010twi}. 
The Pfaffian provides a compact formula for counting the number of perfect matchings in planar graphs—famously applied by Kasteleyn in the dimer model for tiling and matching problems \cite{kasteleyn1961statistics}. 

They are also central tools in random matrix theory, particularly for ensembles with orthogonal and symplectic symmetries (i.e., $\beta=1$ and $\beta=4$ ensembles). Pfaffian structures naturally arise in the calculation of correlation functions, probabilities related to eigenvalue distributions, and averages of characteristic polynomials for these ensembles. Grassmann integrals enable the use of supersymmetry methods to derive explicit formulas for spectral statistics, density of states, and generating functions \cite{efetov2006random,spencer2012susy}. These methods have led to breakthroughs in computing properties such as the smallest eigenvalue distributions in real and quaternion ensembles \cite{akemann2014completing}, and in connecting probabilistic models and integrable systems to random matrix theory via Pfaffian processes and point patterns \cite{ferrari2003random}.

\section{Tensors and Grassmann integrals}
\label{sec:tensors}

\subsection{Hyperpfaffian of an antisymmetric tensor}
Hyperpfaffians have also been introduced in the tensor setting \cite{barvinok1995new} and expressed as a Grassmann integral related to the SYK model 
\cite{Mukhametzhanov_2022} \footnote{The reference \cite{Mukhametzhanov_2022} uses an $i^{p/2}$ in their action that is later compensated by their choice of integration measure.} for fully antisymmetric tensors (otherwise the argument of the exponential vanishes)
\be
{\rm PF}(T)=\int 
D\psi
\exp (
\sum_{1\leq a_1< \dots< a_p\leq N}
T_{a_1\dots a_p} \psi_{a_1}\dots \psi_{a_p})
\label{eq:hyperpf}
\ee
with $p\leq N$, $N$ and $p$ even integers (in order for the tensor, the exponential and the corresponding Hamiltonian  to be Grassmann even) and $p$ dividing $N$.
Allowing for tensors Grassmann odd, our formulas in Section \ref{sec:newdef} 
are also valid for odd integers $p$.
We remark here that our new definition of characteristic polynomials of tensors introduced in Section~\ref{sec:newdef} is related to the hyperpfaffian~\eqref{eq:hyperpf}, see \eqref{eq:pfSquared}.

Unlike the matrix determinant, there are several definitions of hyperdeterminants. The hyperpfaffian \eqref{eq:hyperpf} has been shown \cite{matsumoto2008hyperdeterminantal} closely related 
to the second form introduced by Cayley \cite{cayley}. 
Just as the determinant encodes when a matrix equation has a nontrivial solution and characterizes the singularity of linear maps, the hyperdeterminant identifies the singular locus of multilinear maps and polynomial systems defined by tensors. It serves as a fundamental invariant in algebraic geometry and invariant theory \cite{gelfand2008discriminants}, plays a crucial role in quantum information theory as a measure of entanglement \cite{miyake2003classification}, and provides a critical tool for understanding the solvability and structure of systems of multivariate polynomial equations \cite{ottaviani2012introduction}.

\subsection{
A new definition of 
characteristic polynomials
of a tensor
}
\label{sec:newdef}

Inspired by the rewriting of the matrix characteristic polynomial as \eqref{eq:matrixdet}, we propose to 
generalize this expression to tensors, giving a new notion of 
\emph{tensor characteristic polynomial}.
Our formalism, built on Grassmann integrals, is suitable for 
totally antisymmetric
tensors, a symmetry usually not considered in the literature. We obtain the density of roots of the new characteristic polynomial, analogs of the matrix eigenvalues, and their distribution for Gaussian antisymmetric tensors will lead to similar distributions obtained for the totally symmetric case \cite{gurau2020generalizationwignersemicirclelaw}. Given that such integrals are polynomials
owing to the nature of Grassmann variables,
an $N$ dimensional tensor characteristic polynomial will always have $N$ 
roots,
in contrast with the common exponential in $N$ number of eigenvalues \cite{cartwright2013number}.

Let us take an order-$p$ tensor $T$ of dimension $N$, with 
$p\geq 3$ 
unless specified, 
with
Grassmann variables 
$\{\psi_{a},\psib_a\}_{1\leq a\leq N}$,
and $\lambda$ a complex parameter.
We consider the action to be $\text{SO}(N)$ invariant.
We remark here that $p$ can be either odd or even 
but $T$ must be Grassmann odd or even respectively in order for the action to be Grassmann even.

Consider first $p$ even and the tensor $T$ to be real.
One can write a 
general action
\begin{gather}
S[\lambda,T, \{\psi\},\{\psib\}, \{g\}]
    = 
    \lambda \sum_{a=1}^N \psib_a \psi_a 
    + S_{\text{int}}[T, \{\psi\},\{\psib\}, \{g\}]\\
    S_{\text{int}}[T, \{\psi\},\{\psib\}, \{g\}]=
    \sum_{a_1, \dots, a_p=1}^N 
    \sum_{b_1, \dots, b_p=0}^1
    g^{(b_1 \cdots b_p)}
    \,
    T_{a_1\cdots a_p}
    \prod_{i=1}^p
    \psi_{a_i}^{(b_i)}
    \,,
    \label{eq:actiongeneral}
\end{gather}
with
\begin{equation}
\label{eq:partitionFunctionT}
    Z(\lambda,T, \{g\})
    =\int 
    \mathcal{D}( \psi,\psib) 
    \; 
    e^{S[\lambda,T, \{\psi\},\{\psib\}, \{g\}]}
    \,,
\end{equation}
where 
$\{g^{(b_1 \cdots b_p)}\}$ are any complex constants (including zero),
and
where we defined
\begin{equation}
\psi_{a}^{(0)} := \psi_{a}
\,,
\quad
\psi_{a}^{(1)} := \psib_{a}
    \,,
\end{equation}
and
$\prod_{i=1}^p$
respects the order of $i$ appearing, for example for all $g^{(b_1 \cdots b_p)} = 1$,
{\footnote{
Remark that however, if $g^{(b_1 \cdots b_p)} = 1$ for all $b_i$, then even though a priori, this seems a valid action, one notices after the change of variables to $\psi_+ := \psib + \psi$ and $\psi_- := \psib -\psi$, one can rewrite the part of the action that couples with the tensor to just the product of $\psi_+$. Therefore, this choice of action leads to a trivial one, where the partition function trivially vanishes after integrating over the Grassmann variables.
}}
\begin{align}
\sum_{b_1, \dots, b_p=0}^1
\prod_{i=1}^p 
\psi_{a_i}^{(b_i)}
=
 \psi_{a_1}\cdots \psi_{a_p}  &+\psib_{a_1}\psi_{a_2}\psi_{a_3} \cdots \psi_{a_p}
    + 
    \psi_{a_1}\psib_{a_2}\psi_{a_3} \cdots \psi_{a_p}
    +
    \dots
    + \psi_{a_1}\psi_{a_2} \cdots \psi_{a_{p-1}} \psib_{a_p}
    \nonumber
    \\
     &+\psib_{a_1}\psib_{a_2}\psi_{a_3} \cdots \psi_{a_p}
    + 
    \psib_{a_1}\psi_{a_2}\psib_{a_3} \cdots \psi_{a_p}
    +
    \dots
    + \psib_{a_1}\psi_{a_2} \cdots \psi_{a_{p-1}} \psib_{a_p}
    \nonumber
    \\
    &
    +\dots +\psib_{a_1}\cdots \psib_{a_p}\,.
\end{align}
With this general action above \eqref{eq:actiongeneral},
a priori, $T$ does not assume any symmetry. However, 
depending on the action one chooses to write,
$\psi$ and/or $\psib$ multiplying $T$ projects certain symmetry on the tensor $T$.
This is due to $\psi$ and $\psib$'s Grassmann nature presented in \eqref{eq:grassmannanticommute} 
as well as the fact that all the indices $a_1, \dots, a_p$ are summed over.

For $p$ odd, denoting $\bar{T}$ as another species of Grassmann odd tensor, or for a complex tensor denoting $\bar{T}$ as the complex conjugate of $T$,
\begin{align}
S_{\text{int}}[T, {\bar{T}}, \{\psi\},\{\psib\}, \{g\}, \{\tilde{g}\}]
    = 
    \sum_{a_1, \dots, a_p=1}^N 
    \sum_{b_1, \dots, b_p=0}^1
    \big(
    g^{(b_1 \cdots b_p)}
    \,
    T_{a_1\cdots a_p}
    +
    {\tilde{g}}^{(b_1 \cdots b_p)}
    \,
    {\bar{T}}_{a_1\cdots a_p}
    \big)
    \prod_{i=1}^p
    \psi_{a_i}^{(b_i)}
    \,,
    \label{eq:actiongeneralextra}
\end{align}
with
\begin{equation}
\label{eq:partitionFunctionTextra}
    Z(\lambda,T, {\bar{T}}, \{g\}, \{{\tilde{g}}\})
    =\int 
    \mathcal{D}( \psi,\psib) 
    \; 
    e^{\lambda \sum_{a=1}^N \psib_a \psi_a 
    + S_{\text{int}}[T, {\bar{T}}, \{\psi\},\{\psib\}, \{g\}, \{\tilde{g}\}]}
    \,,
\end{equation}
where $\{\tilde{g}\}$ are another set of any complex constants.

We remark that in order to have a non-trivial $T$ dependence in the partition function $Z(\lambda, T, \{g\})$ (or $Z(\lambda, T)$ in the special case \eqref{eq:partitionFunctionT}),
we need at least $N \ge p$ when $p$ is odd and $N \ge p/2$ when $p$ is even.

Let us consider the matrix case, i.e., $p=2$, for a simple illustration.
Either of the actions below
\begin{gather}
\label{eq:antisymmMaction}
    S_\text{int}[T,\{\psi\},\{\psib\}]
    = 
    \begin{cases}
    \sum_{a_1,  a_2=1}^N T_{a_1 a_2}
    (
    \psib_{a_1}\psi_{a_2}
    +
    \psi_{a_1}\psib_{a_2})\\
    \sum_{a_1,  a_2=1}^N T_{a_1 a_2}
    (
\psi_{a_1} \psi_{a_2} 
    +
    \psib_{a_1}\psib_{a_2}
)
    \end{cases}
\end{gather}
projects $T$ onto its antisymmetric part so that the interaction term is nonzero.
On the other hand, the interaction 
\begin{equation}
\label{eq:symmMaction}
    S_\text{int}[T,\{\psi\},\{\psib\}]
    = 
    \sum_{a_1,  a_2=1}^N T_{a_1 a_2}
    (
    \psib_{a_1}\psi_{a_2}
    -
    \psi_{a_1}\psib_{a_2}
)
\end{equation}
projects
$T$ onto its symmetric part.
We also remark that with $p\geq 3$, the totally symmetric part of the tensor $T$ cannot be extracted in this formulation given in \eqref{eq:actiongeneral} no matter the choice of $\{g_i\}$. This is because each term of $T$ contracting with any product of any combination of more than one $\psi$'s and/or more than one $\psib$'s will necessarily be zero, if $T$ is totally symmetric due to the anticommutation relations \eqref{eq:grassmannanticommute} and the fact that all the indices $a_1, \cdots, a_p$ are summed over.

Let us consider now the special case with $\lambda = 0$ and requiring $g^{(b_1 \cdots b_p)} = 1$ if and only if $b_1 = \cdots = b_p$.
In this case,
the action is given by 
 \footnote{
We need both $\psi$ and $\psib$ so that the partition function which is defined via $\int 
\mathcal{D}( \psi,\psib) 
$ is not zero.}
\begin{equation}
S[T, \{\psi\},\{\psib\}]
    = 
    \sum_{1\leq a_1< \dots< a_p\leq N} T_{a_1\cdots a_p}
    (
    \psi_{a_1}\cdots \psi_{a_p}
    +
    \psib_{a_1}\cdots \psib_{a_p})
    \,,
\end{equation}
then the associated partition function is given by 
\begin{equation}
\label{eq:pfSquared}
    Z(T)
    = \int 
    \mathcal{D}( \psi, \psib) 
    \;  e^{S[T, \{\psi\}, \{\psib\}]}
    =(-1)^{N(N-1)/2}{\text{ PF}}(T)^2
    \,,
\end{equation}
relating to \eqref{eq:hyperpf}.

Because of the Grassmann properties of the variables $\psi,\psib$ 
presented in 
\eqref{eq:grassmannanticommute},\eqref{eq:anticommuteimplication} and \eqref{eq:normalization},
the partition function $Z(\lambda,T, \{g\})$ \eqref{eq:partitionFunctionT} is 
a polynomial of degree $N$ in $\lambda$. 
Then, for a given tensor $T$, the zeros of the partition function \eqref{eq:partitionFunctionT} are ``tensorial" analogs of the matricial eigenvalues determined from \eqref{eq:matrixdet} and the partition function provides a notion of tensor characteristic polynomial that differs from the ones previously introduced in \cite{li2012characteristic, hu2013determinants, galuppi2023characteristic}.
Parallel to the interpretation of the tensor balanced resolvent \cite{gurau2020generalizationwignersemicirclelaw} as a generating function of tensor invariants, the partition function \eqref{eq:partitionFunctionT} can also be seen as a signed generating function of tensor invariants up to order
$\lfloor (2N)/p\rfloor$.

\section{Distributions of the roots of the characteristic polynomials over random Gaussian tensors}
\label{sec:distributionroots}
\begin{figure}
\setlength\tabcolsep{8pt}
\adjustboxset{valign=m}
\begin{tabularx}{\linewidth}{*{3}{>{\centering\arraybackslash}X}}
\adjincludegraphics[width=\linewidth]{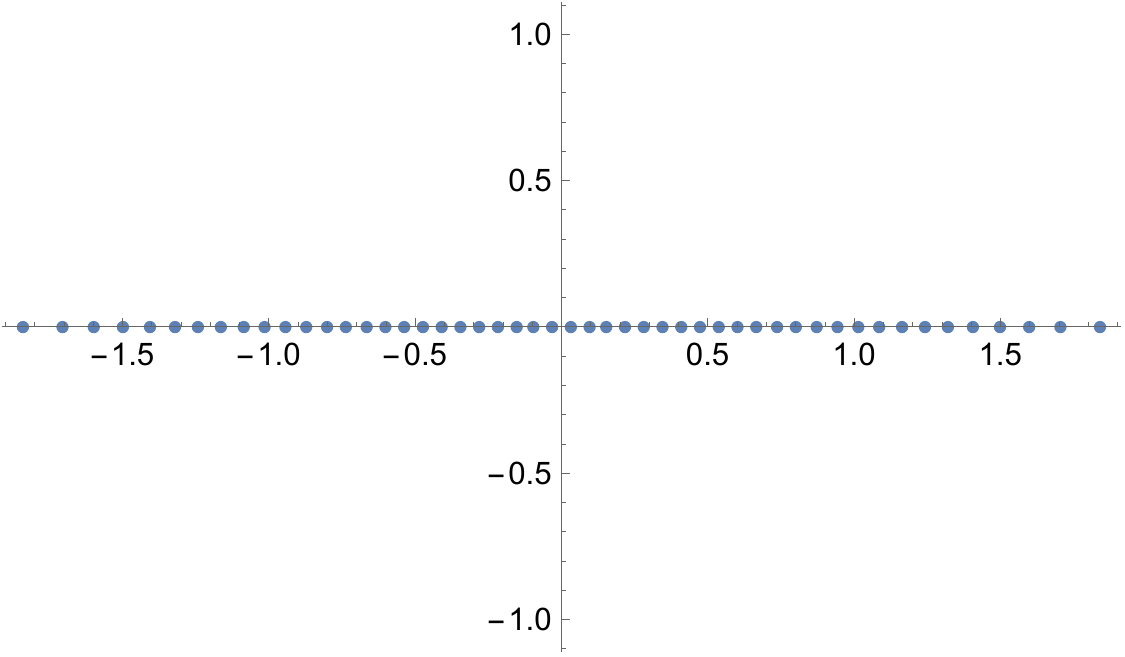}
    &   \adjincludegraphics[height=2cm]{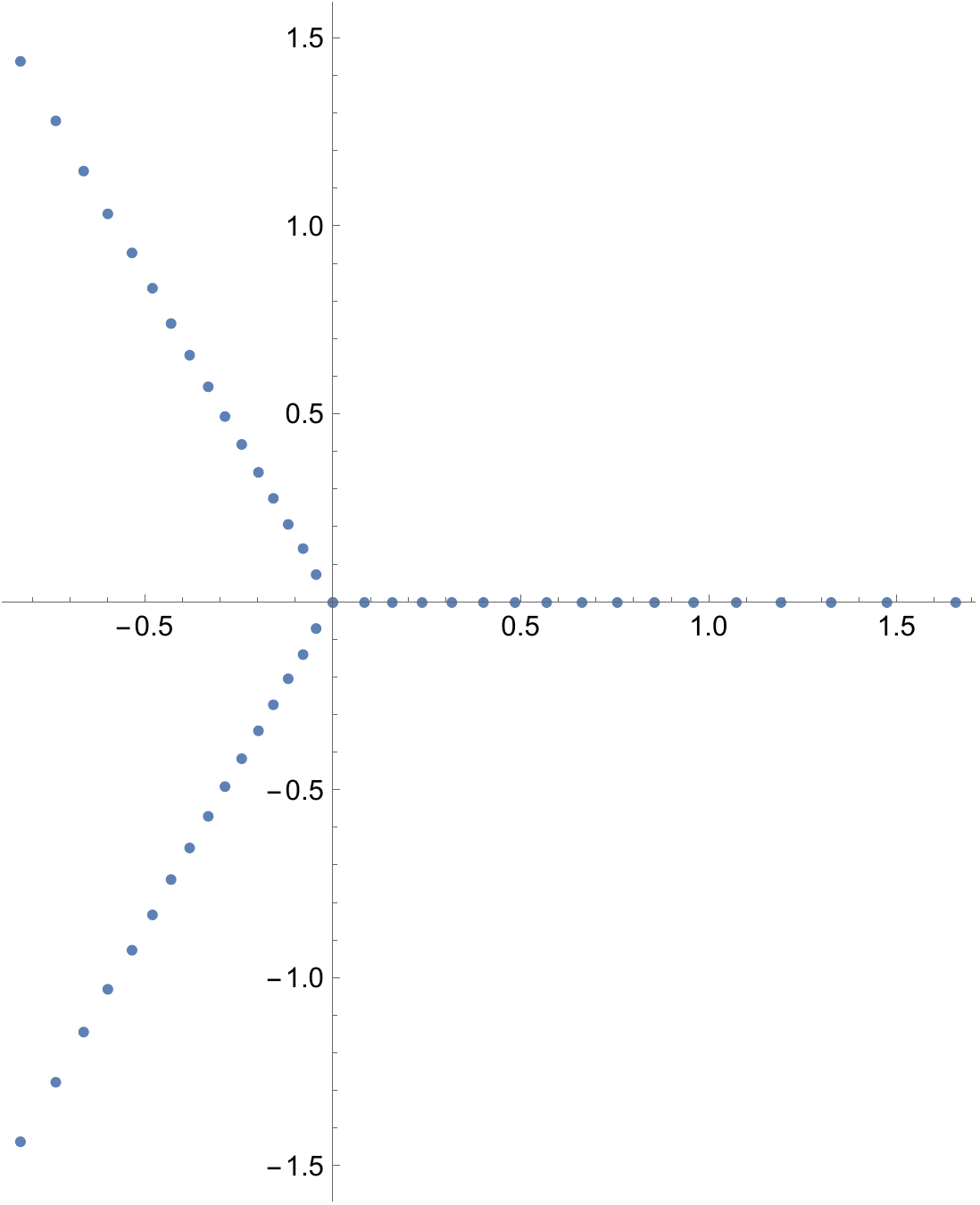}
        &    \adjincludegraphics[height=2cm]{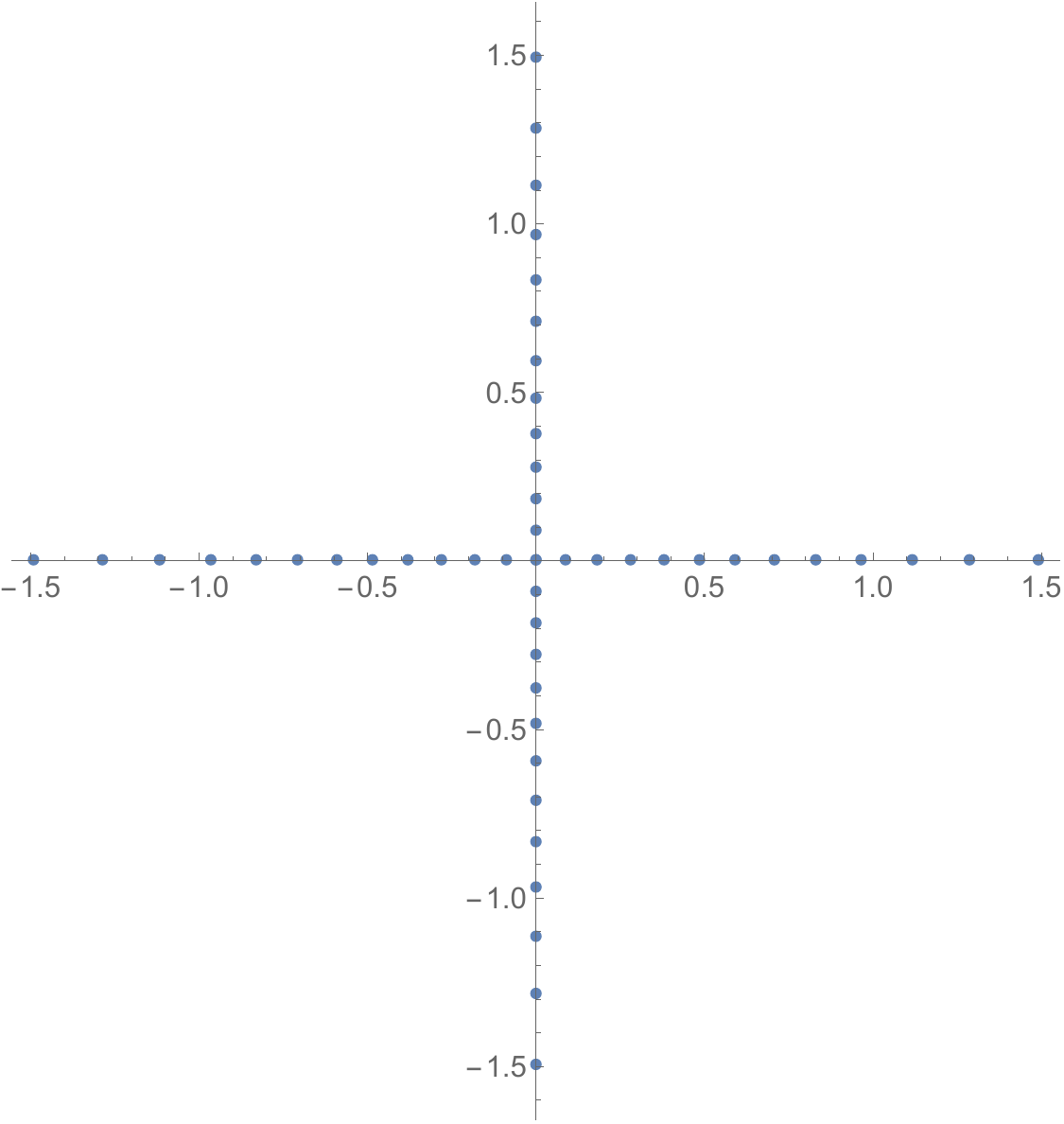}
                \\
\caption*{$p=2$}
    &   
    \caption*{$p=3$}
        &   \caption*{$p=4$}
            \\
\adjincludegraphics[height=2cm]{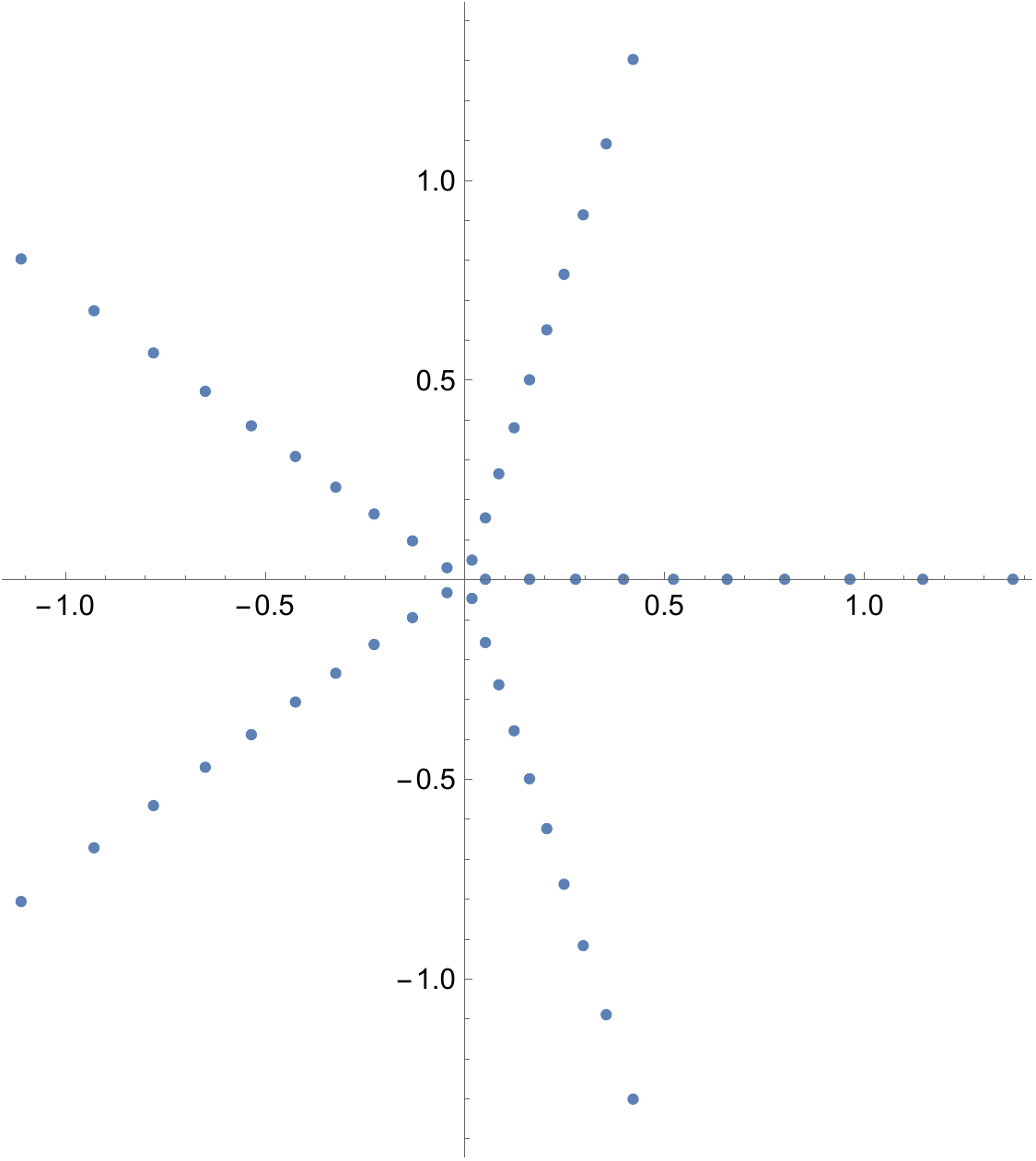}
    &   \adjincludegraphics[height=2cm]{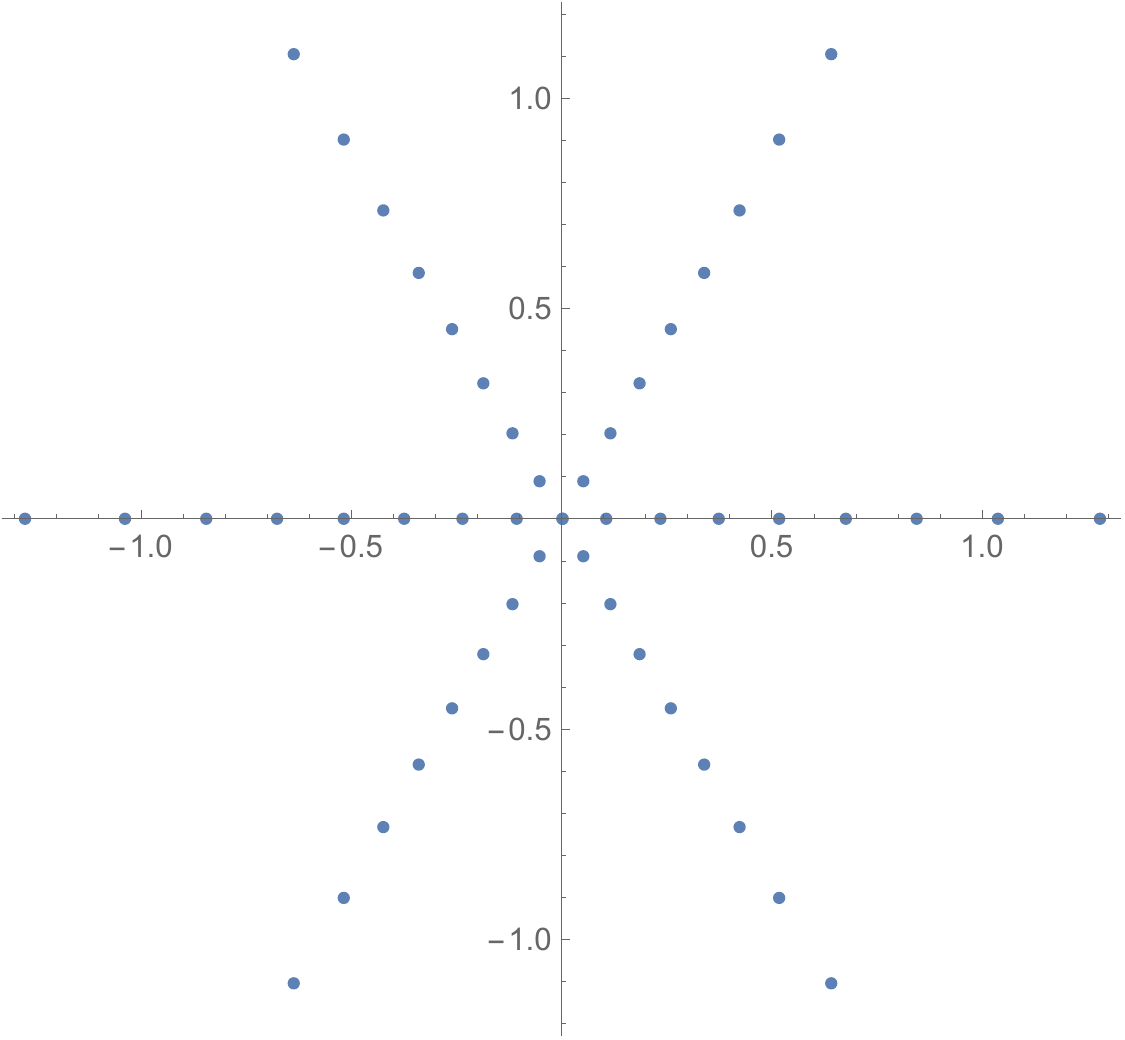}
        &    \adjincludegraphics[height=2cm]{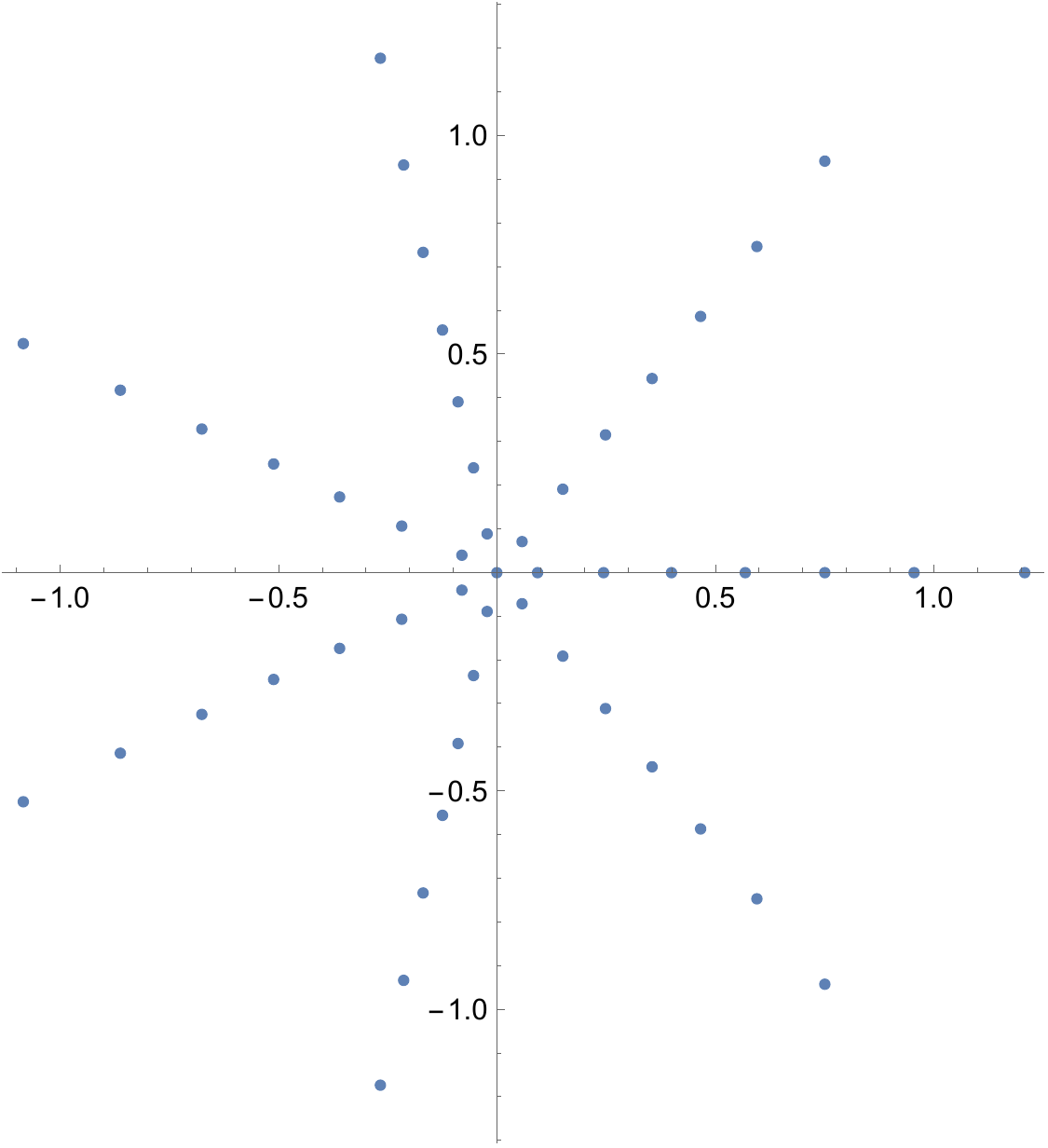}
                \\
\caption*{$p=5$}
    &   
    \caption*{$p=6$}
        &   \caption*{$p=7$}
 \end{tabularx}           
    \caption{Plotting all the zeros of the partition function \eqref{eq:partitionfunctionmonicpolyn} in the complex $\lambda$ plane, with $\mu=N^{1-p}/p$ 
    (i.e., $\tilde{\mu} = 1/p$) for the values $p=2,3,4,5,6,7$ and $N=50$.
    }
    \label{fig:roots}
\end{figure}

Let us now consider random tensor ensembles.
After integrating over 
general real
$T$ drawn from the Gaussian ensemble, 
\begin{equation}
    d \nu (T) = {\mathcal N} \,
    \Bigg[\prod_{a_1, a_2,  \cdots, a_p =1}^N  dT_{a_1 \dots a_p}\Bigg] 
    \exp{- 
N^{p-1}
    \sum^N_{a_1, \dots, a_p = 1} (T_{a_1 \dots a_p})^2}\,,
    \label{eq:measurereal}
\end{equation}
or complex tensor $T$ and its complex conjugate $\bar{T}$, or Grassmann tensors $T$ and $\bar{T}$ of two different species
\begin{equation}
    d \nu (T, \bar{T}) = {\mathcal N} \,
    \Bigg[\prod_{a_1, a_2,  \cdots, a_p =1}^N  dT_{a_1 \dots a_p} d {\bar{T}}_{a_1 \dots a_p}\Bigg] 
    \exp{- 
N^{p-1}
    \sum^N_{a_1, \dots, a_p = 1} {\bar{T}}_{a_1 \dots a_p} \, T_{a_1 \dots a_p}}\,,
    \label{eq:measurecomplex}
\end{equation}
with ${\mathcal N}$ a normalization constant, 
then, regardless of the specific original action that one writes, only certain terms survive in the effective action due to the nature of the Grassmann variables $\psi$ and $\bar{\psi}$. 
Then, we consider the tensor averaged partition function:
\begin{align}
\expval{Z(\lambda,T, \{g\})}_T
\,, \qquad
&{\text{for real tensor $T$,}}
\nonumber
\\
\expval{Z(\lambda,T, {\bar{T}}, \{g\}, \{\tilde{g}\})}_{T, \bar{T}}
\,, \qquad
&{\text{for complex or Grassmann tensors $T$ and $\bar{T}$.}}
\nonumber
\end{align}
In either case, the result takes the general form given below
\begin{align}
    Z(\lambda, \mu)
    &=\int 
    \mathcal{D}( \psi,\psib) 
    \;  \exp (\lambda \sum_{a=1}^N 
\psib_a \psi_a
    -\mu \; \Big(\sum_{a=1}^N
\psib_a \psi_a
    \Big)^p)
    \label{eq:partitionfunctionOurs}
    \\
    =&\lambda^{N} \sum_{n=0}^{\lfloor N/p\rfloor}\frac{1}{n!}\left(-\frac{\mu}{\lambda^p}\right)^n\frac{N!}{(N-pn)!}
    \,,
    \label{eq:partitionfunctionmonicpolyn}
\end{align}
where $\mu$
depends on the original action and is a combination of 
combinatorial factors (coming from the contraction of the Grassmann variables in a way that results into the term 
$\Big(\sum_{a=1}^N\psib_a \psi_a \Big)^p$
) and 
the 
2-point function of the tensor 
\cite{sasakura2023signed}.
We leave $\mu$ general here, but see Appendix \ref{sec:explicitmu} for
some specific examples of actions and corresponding $\mu$.
When $p=2$, the function $Z(\lambda,\mu)$ relates to the Hermite polynomials \cite{delporte2024edgerandomtensoreigenvalues}
as discussed in Section \ref{sec:matrixeigenvaluedet}.
We identify the partition function \eqref{eq:partitionfunctionOurs} expressed as \eqref{eq:partitionfunctionmonicpolyn} with a degree-$N$ monic \footnote{The coefficient of highest degree term is 1.} polynomial in $\lambda$;
therefore, from now on, we drop writing explicitly the $\mu$ dependence from $Z(\lambda, \mu)$ in \eqref{eq:partitionfunctionOurs}, and simply write $Z(\lambda)$.
We can then solve for its roots and some examples are depicted on Fig~\ref{fig:roots}.
In analogy with \eqref{eq:avDetHermite}, the zeros of the averaged characteristic polynomial remind of
eigenvalues of for the tensors.

\subsection{Generating function of the powers of roots} 
The generating function of the sum of the $k$-th powers of the roots $\{\lambda_j^{(N)}\}_{1\leq j\leq N}$ of the monic polynomial 
\be
Z(\lambda) = \sum_{k=0}^N b_k \lambda^k = \prod_{j=1}^N\left(\lambda-\lambda_j^{(N)}\right)
\label{eq:pfmonic}
\ee 
is given by
\be
\mathcal{X}_N(\lambda)=\sum_{k\geq 0} \Xi_N(k)\lambda^k
    \,,
    \quad \Xi_N(k)=\sum_{j=1}^N \Big(\lambda_j^{(N)}\Big)^k
    \,.
\ee
One can show, by expanding explicitly the right hand side below, that 
\begin{equation}
\label{eq:genFunctiontoMonic}
    \mathcal{X}_N(1/\lambda)=\frac{\lambda Z'(\lambda)}{Z(\lambda)}   
    \,,
\end{equation}
where $'$ denotes the derivative with respect to the argument.
In other words, the generating function of the sum of the $k$-th powers of the roots of a monic polynomial can be directly obtained from the polynomial.
We will revisit this expression \eqref{eq:genFunctiontoMonic}  later in Section~\ref{sec:largeNsaddlepoint}, when we present the two-point function for the fermions of the theory that the partition function $Z$ presents.

\subsection{
Large $N$ saddle point analysis and the Fuss-Catalan equation
}
\label{sec:largeNsaddlepoint}
To obtain the large $N$ limit of the partition function, we first have the equality at any $N$ using a ``radial" coordinate $Q$, as done in \cite{gurau2020generalizationwignersemicirclelaw}
\begin{align}
    Z(\lambda)
    &=  \frac{N!}{N^N}\frac{1}{2\pi i}\oint_\mathcal{C} \dd Q \frac{1}{Q^{N+1}}\exp (N\lambda Q - N \tilde{\mu} Q^p)
    \label{eq:actionQ}
\end{align}
introducing 
$\tilde{\mu}=\mu N^{p-1}$
in order to tune the scaling so that we have the desired saddle point analysis as below.
In the second line, we use that the counterclockwise contour integration $\mathcal{C}$ of radius $\epsilon>0$ centered at $Q=0$, is picking up exactly the same contribution, that is the order $N$th term in $\psi \psib$ from the expansion of the exponential, extracted from the Grassmann integral over $\psi$ and $\psib$.

\begin{figure}
    \centering
    \begin{subfigure}{0.3\textwidth}
    \centering
    \includegraphics[width=1\linewidth]{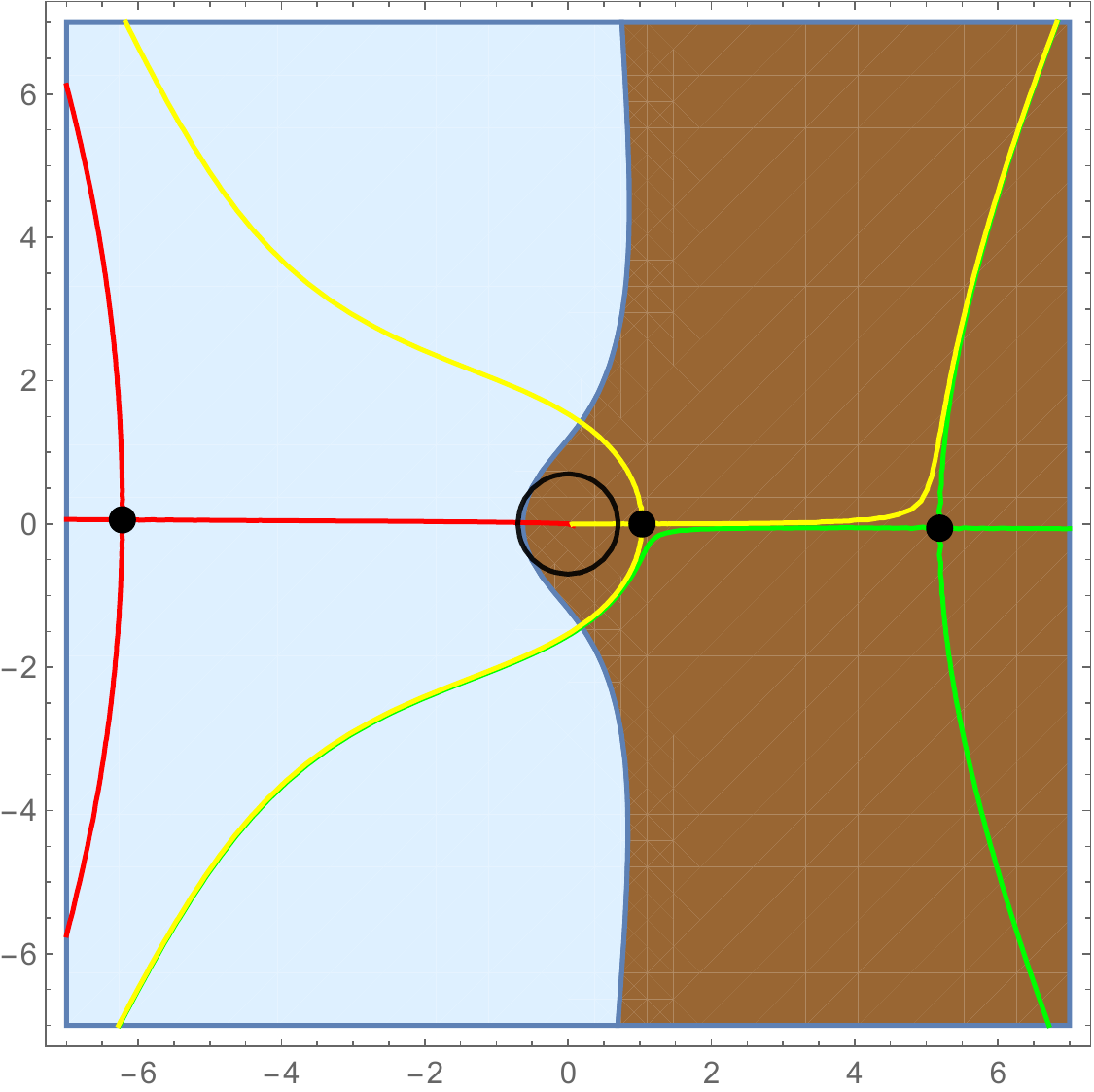}
    \caption{$z_0=0.03 < z_c$, $p=3$.}
    \end{subfigure}
    \begin{subfigure}{0.3\textwidth}
    \centering
    \includegraphics[width=1\linewidth]{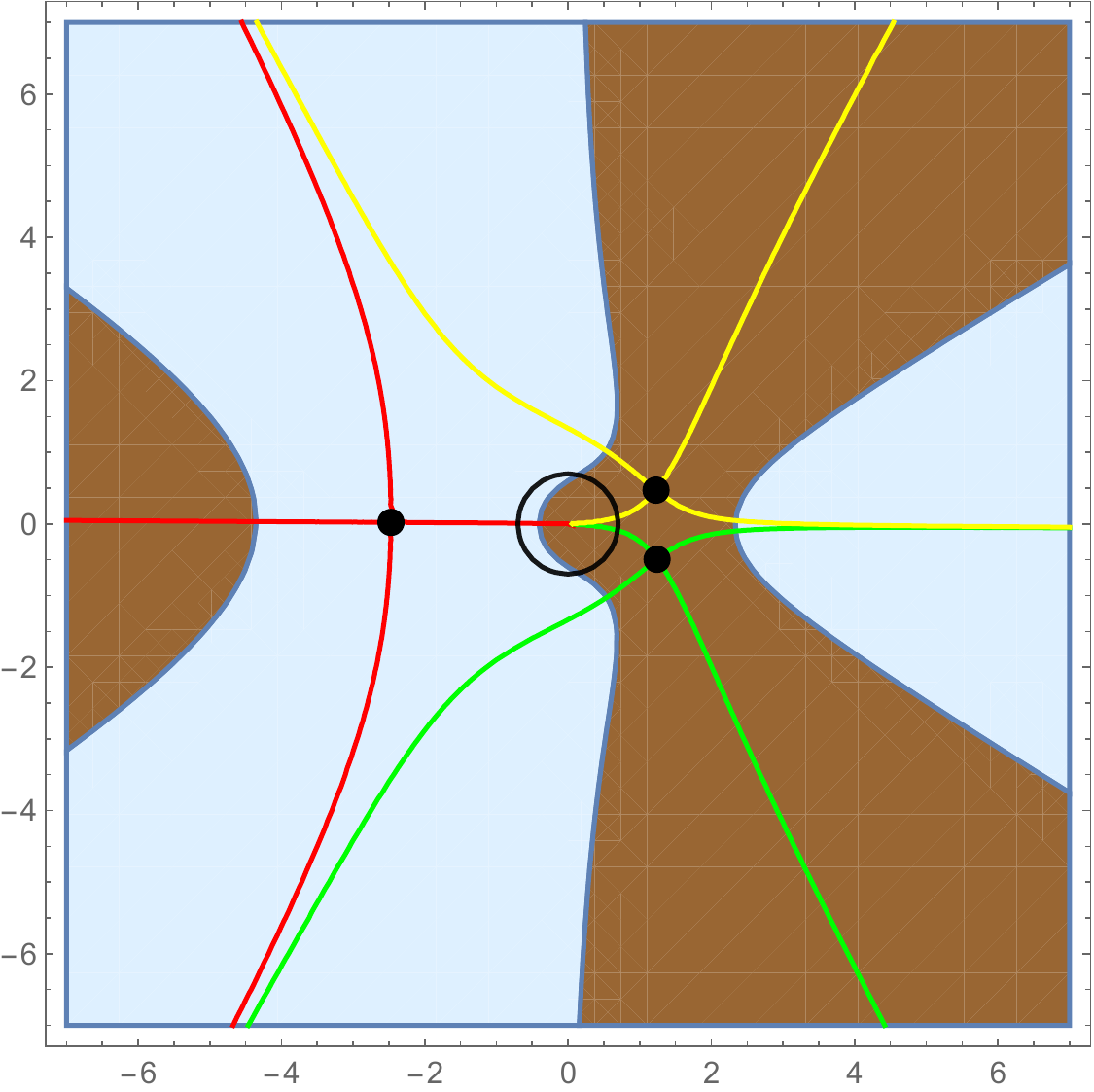}
    \caption{$z_0=0.23 > z_c$, $p=3$.}
    \end{subfigure}\\
    \begin{subfigure}{0.3\textwidth}
    \centering
    \includegraphics[width=1\linewidth]{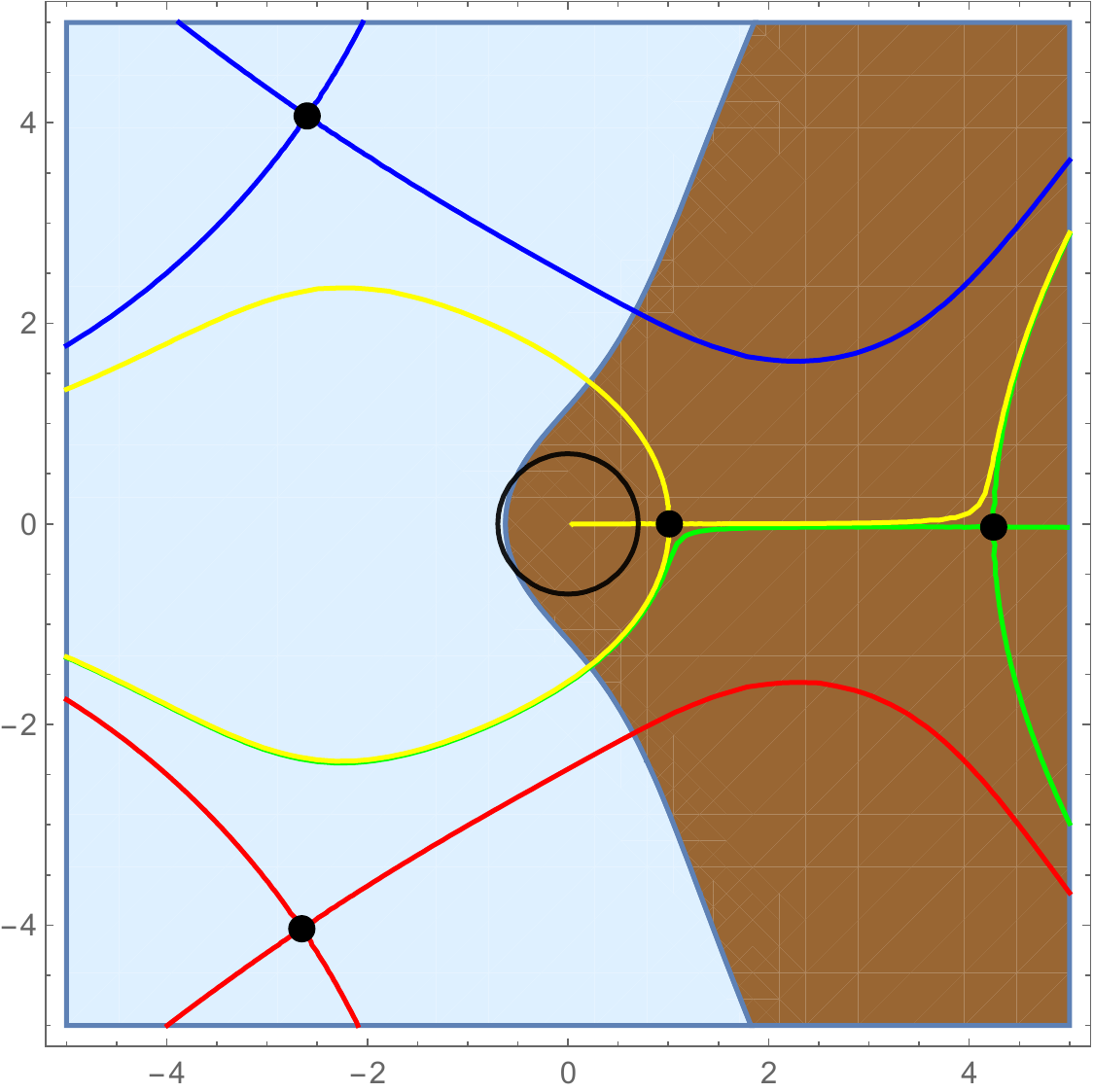}
    \caption{$z_0=0.01 < z_c$, $p=4$.}
    \end{subfigure}
    \begin{subfigure}{0.3\textwidth}
    \centering
    \includegraphics[width=1\linewidth]{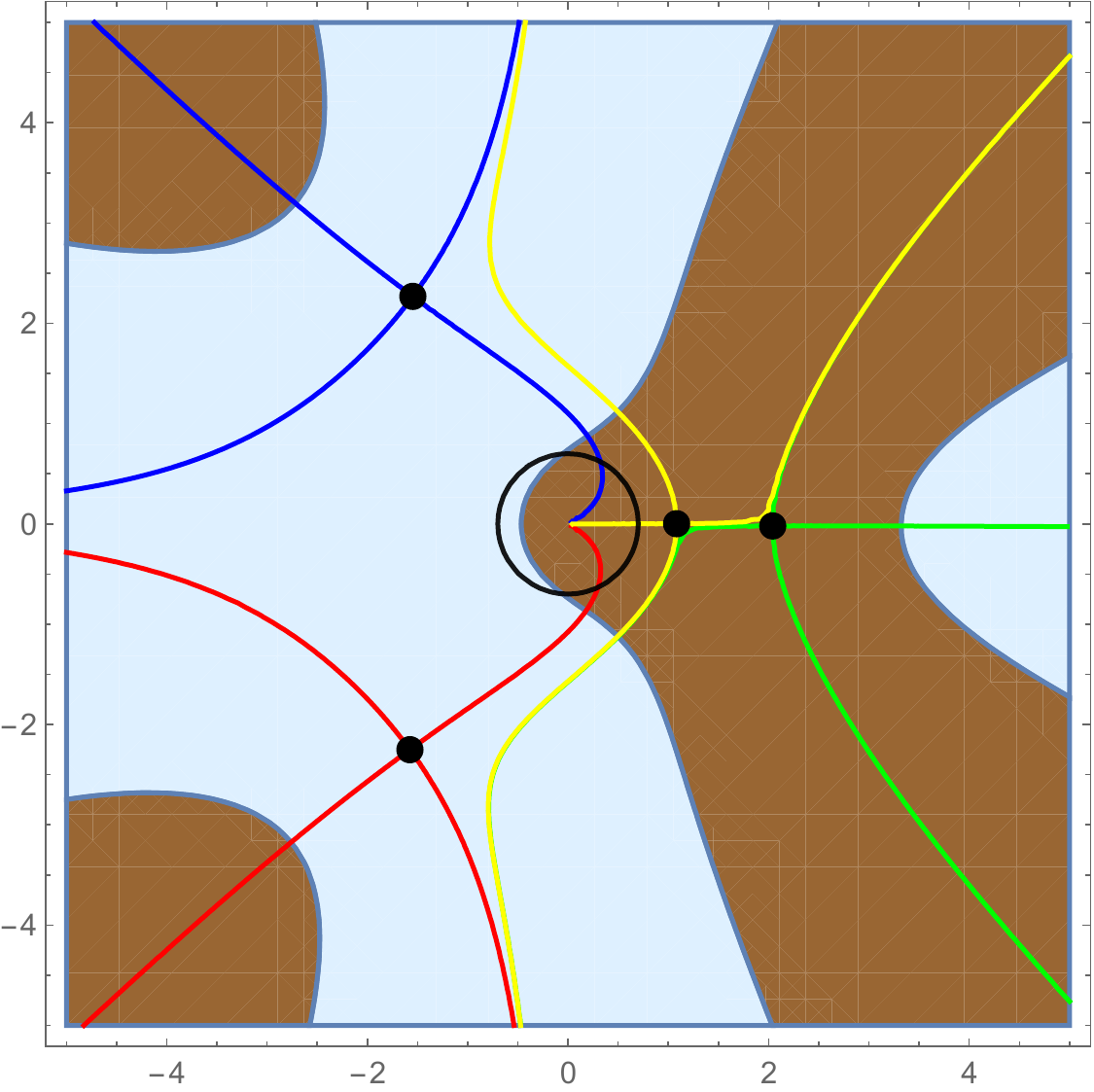}
    \caption{$z_0=0.06 < z_c$, $p=4$.}
    \end{subfigure}
    \begin{subfigure}{0.3\textwidth}
    \centering
    \includegraphics[width=1\linewidth]{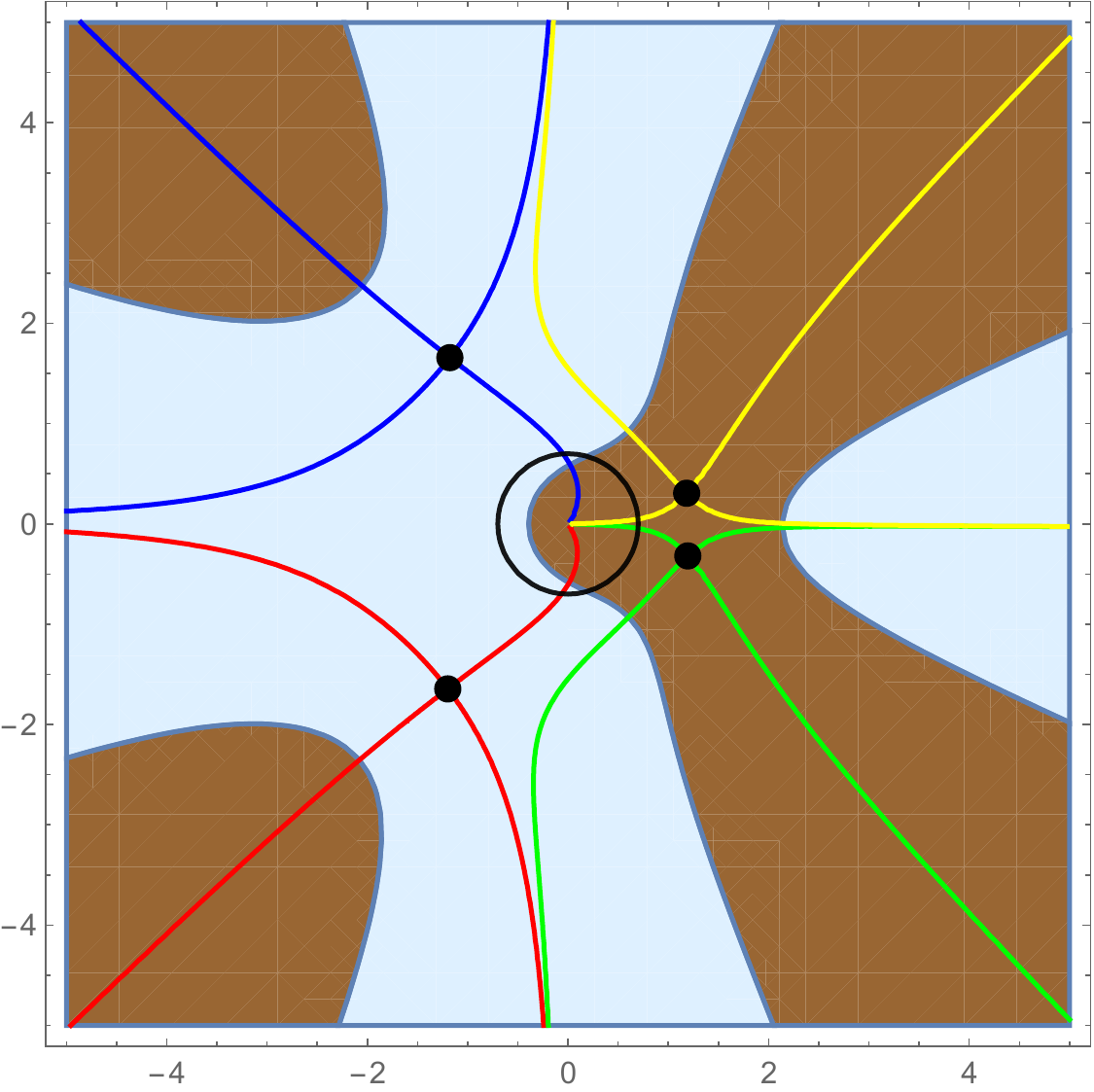}
    \caption{$z_0=0.16 > z_c$, $p=4$.}
    \end{subfigure}
    \caption{We represent in the complex $q$-plane, the Lefschetz thimbles  ending in the light blue regions
    and their duals ending in the brown regions, 
    for each saddle point of the Fuss-Catalan equation, taking the action $S[q]=q-z q^p/p-\log (q/z^{1/p})$.
    The light blue (respectively brown) regions indicate where the real part of the action given in \eqref{eq:largeNactionQ} is negative (respectively positive).
    The black points correspond to the $p$ saddle points given by the $p$ solutions of the Fuss-Catalan equation \eqref{eq:FC}.
    The black circle around the origin is the original curve $\mathcal{C}$ \eqref{eq:actionQ}.
    Only for $z > z_c$, i.e.,(b) and (e), the two saddles of the right for each (yellow and green thimbles and dual thimbles) contribute at leading order in $N$. In the other cases $z < z_c$, i.e., (a) (c) and (d), the saddle point 
    on yellow thimble and dual thimble, contributes at leading order. 
    We have taken $z=z_0e^{i\theta_0}$, $\theta_0=0.02$, with $z_c=2^2/3^3\approx 0.15$ and $z_0=0.03,0.23$  for $p=3$,  and $z_c = 3^3/4^4 \approx 0.10$ and $z_0=0.01,
    0.06,0.16$ for $p=4$.
    }
    \label{fig:lefschetz}
\end{figure}

From \eqref{eq:actionQ},
the large $N$ action and the partition function is given by:
\begin{gather}
    \label{eq:largeNactionQ}
    S[Q]=\lambda Q-\tilde{\mu} Q^p- \log Q \,,\\
    Z\sim\oint_\mathcal{C} \dd Q \exp (NS[Q])\,,
\end{gather}
(where $\sim$ denotes equality at the large $N$ limit and up to a multiplicative factor)
leading to the following saddle point equation:
\begin{gather}
    \lambda Q_*(\lambda)=1+p\tilde{\mu}\left(Q_*(\lambda)\right)^p
\end{gather}
or in the new 
variables 
$q_*=\lambda Q_*$ 
and
$z=\frac{p\tilde{\mu}}{\lambda^p}$,
\begin{gather}
    q_*(z)=1+z q_*(z)^p
    \label{eq:FC}
\end{gather}
that we recognize as the Fuss-Catalan equation, for which $q_0$, one of the $p$ solutions is the generating function of the Fuss-Catalan numbers $F_p(k)$
\begin{gather}
\label{eq:FCgen}
    q_0(z)=\sum_{k\geq 0} F_p(k) z^k\,,\quad F_p(k) =\frac{1}{p k +1}\binom{p k +1}{k}
    \,, 
\end{gather}
around small $z$
with the radius of convergence 
$z_c={(p-1)^{p-1}}/{p^p}$.
Using the Picard-Lefschetz theory \cite{witten2011analytic}, we can see 
which saddles
contribute to the evaluation of the partition function at leading order in $N$ in different regimes. 
The saddle points contribute to the partition function when their dual thimbles (paths ending in the brown regions 
in Fig. \ref{fig:lefschetz})
cross the original contour. For $z<z_c$, among the saddles that indeed contribute, only the saddle \eqref{eq:FCgen} has the largest real value $\Re S(q_0)$ 
(Fig. \ref{fig:lefschetz} (a) for $p=3$ and (c,d) for $p=4$). 
At $z=z_c$, two saddle points collide and for $z>z_c$, those two saddle points have the same real part $\Re S(q_0)$ and opposite imaginary part $\Im S(q_0)$ 
(Fig. \ref{fig:lefschetz} (b) for $p=3$ and (e) for $p=4$). \footnote{We notice multiple Stokes phenomena in the region $z\le z_c$, where Stokes lines join $q_0$ to two other saddles, conjugate of each other, with a negative real value.}

Let us look at the fermionic two-point function: 
\be
\Omega(\lambda):=\frac{\lambda}{N}\dv{}{\lambda}\log Z(\lambda)=\frac{\lambda}{N}\frac{Z'(\lambda)}{Z(\lambda)}
\,,
\ee
which is then identified to be $\mathcal{X}_N(1/\lambda)/N$ 
\eqref{eq:genFunctiontoMonic}.
Using the analysis at the leading order in $N$ when one saddle contributes
\begin{equation}
\frac{1}{N}
\log Z(\lambda) \sim  S[Q_*] 
\,,
\qquad
{\text{with}}
\quad
S[Q_*] =\lambda Q_*-\tilde{\mu} Q_*^p- \log Q_* \,,\\
\quad
{\text{where}}
\quad
\frac{\partial S[Q]}{\partial Q} \Bigg\vert_{Q = Q_*}= 0
\,,
\end{equation}
we compute 
\begin{align} 
    \Omega(\lambda)
    &
    \sim
    \lambda\dv{}{\lambda}
S[Q_*] 
    =
    \lambda \frac{\partial S[Q_*]}{\partial \lambda} 
    +
    \lambda \;
    \frac{\partial S[Q]}{\partial Q} 
    \Bigg\vert_{Q = Q_*}\dv{Q_*}{\lambda}
    \label{omega=Q_*}
    =\lambda Q_* =q_*(z)\,,
    \qquad z\le z_c=\frac{(p-1)^{p-1}}{p^p}
\end{align}
using the Fuss-Catalan equation.
This shows, referring to \eqref{eq:genFunctiontoMonic} that at large $N$, the generating function of the powers of roots of $Z$ is also the generating function of the Fuss-Catalan numbers.
This is in contrast to, but is also a natural generalization of the matrix case which converges to the generating function of the Catalan numbers as discussed in Section \ref{sec:matrixeigenvaluedet}.

\subsection{
Existence and distribution of  
zeros of the partition function in the large $N$ limit
}

We  look for the zeros of the partition function $Z$ in the large $N$ limit. One writes
\begin{align}
Z
\, \sim
\int \dd Q \exp (NS[Q])
\sim \sum_{Q_*} \exp (NS[Q_*])
\,,
\label{eq:saddleZ}
\end{align} 
where $Q_*$ are the saddle points of the action $S[Q]$.
We 
observe that 
the
saddle points
$Q_*$
with the largest value of $\Re S[Q_*]$ will contribute in the large $N$ limit, and the partition function \eqref{eq:saddleZ} has zeros when two saddles 
$Q_*$'s
have the same real part and opposite imaginary parts, for $z>z_c$. 
In that situation, one gets
\begin{align}
Z\sim 
e^{N\Re S[Q_*]}
\;
\cos (N\Im S[Q_*])
\,.
\end{align}
Solving for zeros amounts to ask for 
\begin{equation}
\cos (N\Im S[Q_*])=0 
\; \Leftrightarrow \;
\Im S[Q_*]=
\frac{1}{N}\Big(\frac{\pi}{2}+k \pi\Big)\,,
\qquad k\in \mathbb{Z}\,.
\end{equation}
Noticing that the zeros of $Z$ are located on the interval $z>z_c$ and that they are distributed with a $p$-fold symmetry on the complex $\lambda$ plane, assuming $\tilde{\mu}>0$, we can focus on the distribution of the radial component $r$ of the zeros such that $\lambda=r e^{i 2\pi k/p}$, $0\leq k\leq p-1$. Then, from the equation above, the distance between neighboring zeros is given by
\begin{equation}
\Delta r \dv{\Im S[Q_*]}{r} = \frac{\pi}{N}
\end{equation}
and the density of their absolute value
in the large $N$ limit
is
\begin{gather}
\label{eq:rhoDensity}
\frac{1}{\Delta r}
= \frac{N}{\pi}\dv{\Im S[Q_*]}{r}
= \frac{N}{\pi}\pdv{\Im S[Q_*]}{r}
+ \frac{N}{\pi}
\pdv{\Im S[Q]}{Q}
\Bigg \vert_{Q = Q_*}
\dv{Q_*}{r}\,, 
\end{gather}
using that the saddle points extremize the action, and the normalized spectral density taking into account the $p$-fold symmetry,
\begin{gather}
\label{eq:defrho}
\rho(r)
:=
\frac{p}{ N\abs{\Delta r}}=
\frac{p}{\pi} \abs{\Im Q_*(\lambda)}
\,,
\end{gather}
where the factor $p/N$ has been added to normalize the density $\rho(r)$ to 1 on its domain.

\begin{figure}
    \centering
    \includegraphics[width=0.5\linewidth]{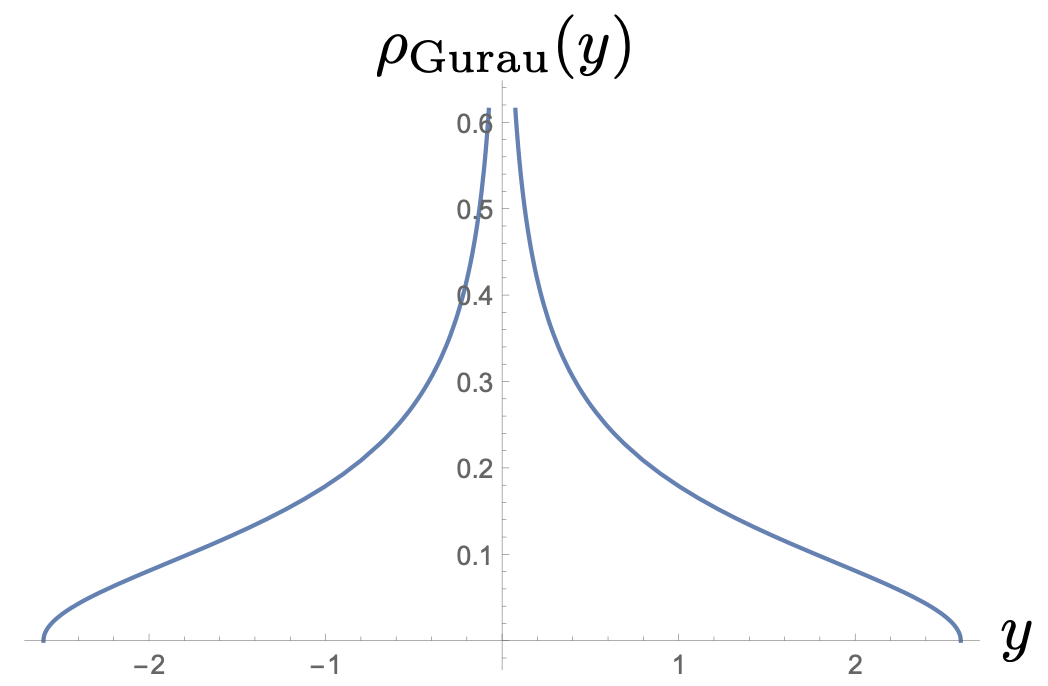}
    \caption{The generalized Wigner law $\rho_\text{Gurau}(y)$ for $p=3$.
    }
    \label{fig:guraup3}
\end{figure}

\paragraph{Gurau's generalized Wigner semi-circle law for real symmetric tensors.}
\cite{gurau2020generalizationwignersemicirclelaw} had introduced a generalized notion of resolvent applied for a 
real
symmetric tensor
$T$ of order $p$
\be
\omega(w)=\frac{1}{w}-\frac{p}{N}\dv{}{w} \expval{\log \mathcal{Z}(w;T)}_T
\label{eq:resolventRazvan}
\ee
with $\mathcal{Z}$ the partition function of the $p$-spin model, 
$w$ is a complex coupling constant
\begin{equation}
    \mathcal{Z}(w;T)=\int_{\mathbb{R}^N} D\phi \,  \exp (-\frac{1}{2}\phi^2 + \frac{1}{w \,  p}T\cdot \phi^p)\,,
    \qquad
   D \phi =  (2 \pi)^{-N/2} \prod_{i = 1}^N d \phi_i
    \label{eq:partitionfunctionRazvan}
\end{equation}
where the average $\expval{\cdot}_T$ was taken over $T$ drawn from the Gaussian ensemble given in
\begin{equation}
    d \nu (T) = Q_N \,
    \Bigg[\prod_{a_1 \le \cdots \le a_p}  dT_{a_1 \dots a_p}\Bigg] 
    \exp{- \frac{N^{p-1}}{2p}\sum^N_{a_1, \dots, a_p = 1} (T_{a_1 \dots a_p})^2}\,,
\end{equation}
with a normalization constant $Q_N$.
\footnote{Because the partition function at given tensor $T$ possesses two infinite cuts from 0 to $\pm \infty$ in the $w$ plain, one needs to give a small imaginary component for the $\phi$ integral determining its analytic continuation. The jump across the cuts is given by a sum over instanton contributions, in one to one correspondence with (unnormalized) E-eigenpairs, in the sense of Def.~1.1 of \cite{cartwright2013number}.}
At large $N$, relying on an annealed evaluation of the two-point function of the Gaussian $p$-spin model, the resolvent $\omega$ was given by 
\be
\label{eq:gurauResolvent}
\omega(w)
=\frac{1}{w}q_0 \Big(\frac{1}{w^2}\Big) 
=\int_{-w_c}^{w_c} \dd y\frac{w \abs{y}P_p(y^2)}{w^2-y^2}
\,,
\ee 
with explicit expressions for $p=2,3$ given in \cite{gurau2020generalizationwignersemicirclelaw}, leading to associated 
normalized spectral densities 
($\int \rho_{\text{Gurau}} (y) \dd y
=1$)
evaluating $\omega$ across its cut on $[-w_c,w_c]$ 
via the Sokhotski-Plemelj formula
\begin{gather}
    \rho_{\text{Gurau}}(y)=\frac{1}{2 \pi i}\lim_{\epsilon \to 0} \left( \omega(y - i \epsilon) - \omega (y + i \epsilon)\right)= \abs{y} P_p(y^2) \,,
    \\ 
    y \in (-w_c,w_c)\,,
    \qquad w^2_c=\frac{p^p}{(p-1)^{p-1}}
    \,,
\end{gather}
and 
$P_p$ 
is given in \eqref{eq:FCs1}, App.~\ref{app:FC}. A plot of that spectral density for $p=3$ is shown on Fig.~\ref{fig:guraup3}.

\paragraph{Density distribution of zeros.}
Comparing our result \eqref{omega=Q_*}, \eqref{eq:FC}, \eqref{eq:FCgen} with Gurau's result
\eqref{eq:gurauResolvent} for the 2-point functions, we identify that  $w^2 = z^{-1}$, therefore
\begin{equation}
w^2 = \frac{1}{p \, \tilde{\mu}} \lambda^p
\,.
\label{eq:wz}
\end{equation}
We have the equivalence of two probability distributions of total mass one,
we have
\begin{equation}
\frac{1}{2}\rho (r) \dd r
=
\rho_{\text{Gurau}} (y) \theta(y)\dd y
\label{eq:equalprobdistr}
\end{equation} 
with $\theta(\cdot)$ the Heaviside function, such that $\rho_{\text{Gurau}} (y)$ ($y \in (-w_c,w_c)$) and $\rho (r)$ ($r \in [0,w_c^{1/p}]$) are both probability distributions on their respective domain.
\footnote{Remarking that the effective actions in radial coordinates $Q$ and $\rho$ (eq. (37) in \cite{gurau2020generalizationwignersemicirclelaw}, not to be confused with our $\rho$, eq.~\eqref{eq:defrho}) are analogous (up to a $\log \lambda$ term and a global 
$-1/2$
factor, that do not change the saddle point equation), using 
\eqref{eq:wz}
and $\rho^2=\lambda Q$.}
One then recovers the distributions of \cite{gurau2020generalizationwignersemicirclelaw} with the change of variable 
$y=r^{p/2}/\sqrt{p\tilde{\mu}}$ 
in the results of Section 4.1 of \cite{gurau2020generalizationwignersemicirclelaw}
\begin{equation}
\label{eq:TensorWignerLaw}
    \rho(r)=
    \left(\sqrt{\frac{p}{\tilde{\mu}}} \, r^{p/2-1}\right)\rho_{\text{Gurau}}\left(\frac{r^{p/2}}{\sqrt{p\tilde{\mu}}}\right)
    \,.
\end{equation}

\begin{figure}
    \centering
    \includegraphics[width=0.5\linewidth]{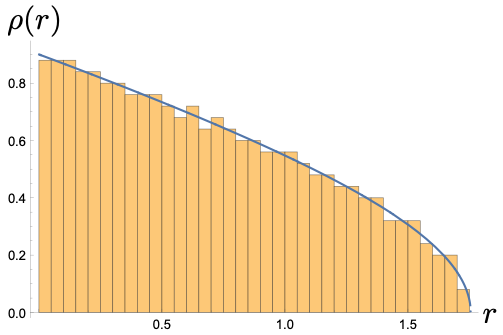}
    \caption{The distribution of the absolute value $r$ of the roots from \eqref{eq:TensorWignerLaw} superimposed with the histogram of the absolute value of the roots of the polynomial \eqref{eq:partitionfunctionOurs}, for $N=2000$, $p=4$ and $\tilde{\mu}=1/p$.
    }
    \label{fig:placeholder}
\end{figure}

The reference \cite{gurau2020generalizationwignersemicirclelaw} developed a notion of eigenvalue adapted to tensors that emerges from the analytic structure of a bosonic partition function, in particular, the discontinuity at its branch cuts. 
In contrast, our fermionic partition function is a polynomial at finite $N$. The generating function $\mathcal{X}$ of the sums of powers of zeros obeys the same large $N$ equation as the two-point function of the partition function of Gurau. 
The consequence of this equality at large $N$ is that the two distributions derived on the one side from the generalized resolvent, and on the other from the generating function $\mathcal{X}$ are the same up to an explicit and simple change of variables.

\section{Concluding remarks}
\label{sec:conclusion}
Our formulation of a tensor characteristic polynomial in terms of Grassmann variables provides a novel interpretation for a tensor spectral density leading to Fuss-Catalan distributions and we relate it to the generalized Wigner semi-circle law proposed by \cite{gurau2020generalizationwignersemicirclelaw}. It is suitable for tensor symmetries other than the fully symmetric one, in particular the totally antisymmetric. It is also valid at finite $N$, the number of associated roots of the generalized characteristic polynomial (always $N$ solutions) differing from the number of saddle points of the spherical $p$-spin model (at most exponential in $N$) 
\cite{cartwright2013number,auffinger2013random}. However such spectrum will be associated to Grassmann eigenvectors, fitting the framework of exterior algebras. It would also be interesting to relate the partition formulas obtained here for tensors of order $p$ to the finite $N$ results of the product of $p-1$ Ginibre matrices \cite{penson2011product}.
Another open question is the comparison of the quenched partition function to the annealed one in the light of Ref.~\cite{baldwin2020quenched}. We leave those explorations for future works.

\section*{Acknowledgements}
NS is supported in part by JSPS KAKENHI Grant No.~25K07153. ND thanks Cihan~Pazarba\c{s}{\i} for helpful discussions and for sharing his code on Lefschetz thimbles.
RT thanks the discussion with Juan~Abranches.

\newpage
\appendix

\section{Edwards-Jones formula}
\label{app:EJformula}

We first recall the Sokhotski–Plemelj theorem
\begin{equation}
    \lim_{\epsilon\to 0} \frac{1}{x\pm i \epsilon}=\text{Pr}\left(\frac{1}{x}\right) \mp i \pi\delta(x)
\end{equation}
where $\text{Pr}(\cdot)$ denotes the principal part of $(\cdot)$. This allows to rewrite the spectral density of a given matrix $M$ with eigenvalues $\{\lambda_i\}_{1\leq i\leq N}$:
\begin{align}
    \rho_N(\lambda)
    &=\frac{1}{N}\sum_i\delta(\lambda-\lambda_i)\\
    &=\frac{1}{\pi N}\lim_{\epsilon\to 0} \Im \sum_i \frac{1}{\lambda - \lambda_i - i \epsilon}\\
    &=\frac{1}{\pi N}\lim_{\epsilon\to 0} \Im \sum_i \dv{}{\lambda}\log(\lambda-\lambda_i-i\epsilon)\\
    &=\frac{1}{\pi N}\lim_{\epsilon\to 0} \Im  \dv{}{\lambda}\log \det(\lambda-M-i\epsilon)
\end{align}
where in the second line, the $\Im$ gets rid of the principal part. In the third line, we used the principal branch of the complex logarithm and in the fourth $\Tr \log (\cdot)= \log \det(\cdot)$.

\section{Fuss-Catalan distributions
}
\label{app:FC}

Based on \cite{penson2011product}, one has for 
$k\in \mathbb{N}$ 
and 
$p\in \mathbb{N}\setminus \{0, \; 1\}$
that the Fuss-Catalan numbers \footnote{In the paper \cite{penson2011product}, they use $FC_s(k)$ that corresponds to $F_{s+1}(k)$ in our notation. For example, with $p=2$ or $s=1$, we obtain the Catalan numbers.}
\begin{equation}
F_p(k) := \frac{1}{p k + 1} \binom{p k  + 1}{k}
\end{equation}
are associated to the density distribution $P_p(x)$ 
which can be formally written as an inverse Mellin transform
\begin{gather}
    F_p(k)=\int_0^{1/z_c}\dd x ~ x^k P_p(x)
    \,, 
    \qquad 
    P_p(x)= \mathcal{M}^{-1}[F_p(\sigma);x]
    \,,
    \qquad
    z_c=\frac{(p-1)^{p-1}}{p^p}
    \,.
\end{gather}
and $P_p$ explicitly written as 
\begin{equation}
\label{eq:FCs1}
   P_p(x)  = \sum_{n=1}^{p-1} \Lambda_{n,p} \;
     x^{\frac{n -p }{p}}  {}_{p-1}F_{p-2}
     \bigg(  \bigg\{ 1 - \frac{1+m}{p-1} +\frac{n}{p}\bigg\}_{m=1}^{p-1}  ,  
     \bigg\{ 1 + \frac{n-m}{ p }\bigg\}_{\genfrac{}{}{0pt}{}{m=1}{ m\neq n } }^{p-1}   ; \frac{(p-1)^{p-1}}{p^p}x\bigg) \;,
\end{equation}
where the coefficients $\Lambda_{n,p}$ read for $n=1,2, \dots, p-1$
\begin{equation}
\label{eq:FCs2}
 \Lambda_{n,p} = \frac{1}{(p-1)^{3/2}} \sqrt{\frac{p}{2\pi}}
 \left( \frac{ (p-1)^{p-1 }}{p^p}\right)^{\frac{ n}{p} }
 \;   \frac{ 
   \prod_{m=1,\dots, p-1}^{m\neq n } \Gamma\left( \frac{m-n}{p} \right) }
  { \prod_{m=1}^{p-1} \Gamma \left( \frac{m+1}{p-1} -\frac{n}{p}  \right) } \;,
\end{equation}
while $\Gamma(\cdot)$ is the Gamma function and  ${}_pF_{p-1}(\cdot, \cdot, \cdot, \cdot)$ is the hypergeometric function.

\section{Details on the choice of actions}
\label{sec:explicitmu}
We will give an explicit computation of the effective couplings $\mu$ appearing in \eqref{eq:partitionfunctionOurs} for different interaction terms. Let us take for simplicity a totally antisymmetric tensor $T$ with the two-point function averaged over random Gaussian tensor ensembles,
\begin{align}
    \expval{T_{a_1\cdots a_p}T_{a_1\cdots a_p}}&=\beta/N^{p-1}\quad \text{for $p$ even and $T$ real}\,,
        \quad 
    {\text{with}} \; 
    \expval{\cdot}:=\int \dd \nu(T) \,, 
    \;
    {\text{or}}
    \label{eq:2pttensorreal}
    \\
    \expval{T_{a_1\cdots a_p}\bar{T}_{a_1\cdots a_p}}&=\beta/N^{p-1}\quad \text{for $p$ odd or $T$ complex}\,, 
    \quad 
    {\text{with}} \;
    \expval{\cdot}:=\int \dd \nu(T,\bar{T}) \,.
    \label{eq:2pttensorcomplex}
\end{align}
without summation and imposing $1\leq a_1<\cdots<a_p\leq N$, 
and where $\beta = 1/2$ for \eqref{eq:2pttensorreal} and $\beta = 1$ for \eqref{eq:2pttensorcomplex} if the Gaussian probability measures were given in \eqref{eq:measurereal} and \eqref{eq:measurecomplex} respectively, restricting to indices of strictly increasing order.
\footnote{If one defines this tensor $T$ as the totally antisymmetric part of a tensor $\tilde{T}$
\begin{equation}
    T_{a_1\cdots a_p}=\frac{1}{p!}\sum_{\sigma\in S_p}\text{sign}(\sigma)\tilde{T}_{\sigma(a_1)\cdots \sigma(a_p)}\,, \quad 1\leq a_1<\cdots<a_p\leq N\,,
\end{equation}
for which the two-point correlation is given by
\begin{equation}
    \expval{\tilde{T}_{a_1\cdots a_p}\tilde{T}_{a'_1\cdots a'_p}}=\gamma^2\,\delta_{a_1 a'_1}\dots \delta_{a_p a'_p}\,, \quad 1\leq a_1,\dots, a_p\leq N\,,
\end{equation}
with a constant $\gamma$, then the two-point function of $T$ is given by 
\begin{equation}
    \expval{T_{a_1\cdots a_p}T_{a'_1\cdots a'_p}}=\frac{\gamma^2}{p!}\,\delta_{a_1 a'_1}\dots \delta_{a_p a'_p}\,,\quad 1\leq a_1<\cdots<a_p\leq N\,.
\end{equation}}

One can concisely write the interactions that will be considered as follows:
\begin{gather}
    S_{\text{int}}[T, \{\psi\},\{\psib\}]
    = \alpha \sum_{1\leq a_1< \dots< a_p\leq N} T_{a_1\cdots a_p}(J_{a_1\cdots a_p}+\bar{J}_{a_1\cdots a_p})
    \,,
\end{gather}
with $p$ even and $T$ real, or 
\begin{align}
S_{\text{int}}[T,\bar{T}, \{\psi\},\{\psib\}]
    =  
    \alpha \sum_{1\leq a_1< \dots< a_p\leq N} \left(\bar{T}_{a_1\cdots a_p}J_{a_1\cdots a_p}+\bar{J}_{a_1\cdots a_p}T_{a_1\cdots a_p}\right)
    \,,
\end{align}
for $p$ odd or $T$ complex, 
with $J_{a_1\cdots a_p}$ and $\bar{J}_{a_1\cdots a_p}$ functions of $\psi$ and $\psib$ and $\alpha$ a coupling constant.
\begin{itemize}
    \item Taking first:
     \begin{equation}
    J_{a_1\cdots a_p}=\psi_{a_1}\cdots \psi_{a_p}
    \,,
     \qquad
     \bar{J}_{a_1\cdots a_p}=\psib_{a_1}\cdots \psib_{a_p}
     \end{equation}
the effective action after the average over $T$ (and $\bar{T}$ for $p$ odd or a complex tensor) is
\begin{gather}
S[\lambda, \{\psi\},\{\psib\}]
    = 
    \lambda \psib\cdot \psi
    -\mu (\psib\cdot \psi)^p
    \,,
     \label{eq:effaction1}
    \\
    \mu=\frac{\alpha^2 \beta \zeta}{p!N^{p-1}}\,,\quad 
    \zeta=(-1)^{\frac{p(p- 1)}{2}-1}\,,
    \nonumber
\end{gather}
where
we have written $\psib\cdot \psi:=\sum_{a=1}^N \psib_a \psi_a$. \footnote{Strictly speaking, the parities of $p(p-1)/2$ and $p(p+1)/2$ are the same when $p$ is even and only the first is present when $p$ is odd with our way of writing the action.\label{foot:oddvseven}}
In order to show \eqref{eq:effaction1}, let us start with $p$ even and $T$ real. 
After averaging over $T$, we obtain
\begin{equation}
    \frac{\alpha^2 \beta }{2 N^{p-1}}\sum_{1\leq a_1< \dots< a_p\leq N}(J_{a_1\cdots a_p}\bar{J}_{a_1\cdots a_p}+\bar{J}_{a_1\cdots a_p}J_{a_1\cdots a_p})
    \,.
    \label{eq:averageTreal}
\end{equation}
The first term corresponds to
\begin{align}
    J_{a_1\cdots a_p}\bar{J}_{a_1\cdots a_p}
    =
    \psi_{a_1}\cdots\psi_{a_p}\psib_{a_1}\cdots\psib_{a_p}
    =
    (\psib_{a_1}\psi_{a_1})\cdots(\psib_{a_p}\psi_{a_p})(-1)^{\frac{p(p+1)}{2}}
    \,,
\end{align}
where the last equality is obtained by permuting all the $\psib$'s to bring them next to their corresponding $\psi$, collecting as many minus signs as:
\begin{equation}
    p+(p-1)+\cdots+1=\frac{p(p+1)}{2}
    \,.
\end{equation}
The $\bar{J}J$ term in \eqref{eq:averageTreal} gives a similar contribution
\begin{align}
    \bar{J}_{a_1\cdots a_p}J_{a_1\cdots a_p}
    =
    \psib_{a_1}\cdots\psib_{a_p}\psi_{a_1}\cdots\psi_{a_p}
    =
    (\psib_{a_1}\psi_{a_1})\cdots(\psib_{a_p}\psi_{a_p})(-1)^{\frac{p(p-1)}{2}}\,,
\end{align}
effectively cancelling the $1/2$ factor in \eqref{eq:averageTreal}.
Additionally, we have
\begin{equation}
    \sum_{1\leq a_1<\dots<a_p\leq N}(\psib_{a_1}\psi_{a_1})\cdots(\psib_{a_p}\psi_{a_p})=\frac{1}{p!}\left(\psib\cdot \psi\right)^p
    \,.
\end{equation}
Combining all the factors, 
we arrive at \eqref{eq:effaction1}.

For $p$ odd or $T,\bar{T}$ complex, we recall this Gaussian integration formula
\begin{equation}
    \int \mathcal{D}(T,\bar{T}) \exp(-\bar{T}\cdot T+\bar{T}\cdot J + \bar{J}\cdot T)=\exp(\bar{J}\cdot J)
\end{equation}
such that the average over $T$ gives
\begin{equation}
    \frac{\alpha^2 \beta }{N^{p-1}}\sum_{1\leq a_1< \dots< a_p\leq N}\bar{J}_{a_1\cdots a_p}J_{a_1\cdots a_p}
\end{equation}
and that is enough to conclude. 
\item Similarly, taking as interaction terms
    \begin{gather}
        J_{a_1\cdots a_p}=
            \psi_{a_1}\cdots \psi_{a_k}\psib_{a_{k+1}}\cdots \psib_{a_p}\\
        \bar{J}_{a_1\cdots a_p}=
        \psib_{a_1}\cdots \psib_{a_{k}}\psi_{a_{k+1}}\cdots \psi_{a_p}
    \end{gather}
for an integer $k$ ($1\leq k<p$), we obtain the effective action
\begin{gather}
    S[\lambda, \{\psi\},\{\psib\}]
    = 
    \lambda \psib\cdot \psi
    -\mu(\psib\cdot \psi)^p\,,\label{eq:effaction3}\\
    \mu=
    (-1)^{p(p-1)/2+(p-k)-1}\frac{\alpha^2 \beta}{p!N^{p-1}} 
    \,.\notag
\end{gather}
    \item Let us now consider the following interaction: 
    \begin{gather}
    J_{a_1\cdots a_p}=\psi_{a_1}\cdots \psi_{a_{p-1}}\psib_{a_p}+\cdots+\psib_{a_1}\psi_{a_2}\cdots \psi_{a_p}\,,
    \\ \bar{J}_{a_1\cdots a_p}=\psib_{a_1}\cdots \psib_{a_{p-1}}\psi_{a_p}+\cdots+\psi_{a_1}\psib_{a_2}\cdots \psib_{a_p}\,,
\end{gather}
where there is a single factor $\psib$ for each term in $J$ and in the second line, there is a single factor $\psi$ in $\bar{J}$. Then, the effective action after the average over $T$ (and $\bar{T}$ for $p$ odd or a complex tensor) 
is
\begin{gather}
    S[\lambda, \{\psi\},\{\psib\}]
    = 
    \lambda \psib\cdot \psi
    -\mu (\psib\cdot \psi)^p
    \,,
    \nonumber
    \\
    \mu = \frac{\alpha^2 \beta  \zeta p}{p!N^{p-1}}
    \,,\quad 
    \zeta=(-1)^{\frac{p(p- 1)}{2}}\,,
    \label{eq:effaction2}
\end{gather}
where the same comment as in \cref{foot:oddvseven} applies.
Now, for the demonstration purpose, we will focus on the real $T$ case only.
We recall first the notation for totally antisymmetrized set of indices:
\begin{equation}
    J_{[a_1\cdots a_p]}:=\frac{1}{p!}\sum_{\sigma\in S_p} \text{sign}(\sigma)J_{\sigma(a_1)\cdots\sigma(a_p)}
    \,.
\end{equation}
Then, doing the analysis for $p$ even, but it can be repeated for $p$ odd and $T$ complex, the interaction term can be rewritten as:
\begin{gather}
    \alpha \sum_{1\leq a_1< \dots< a_p\leq N} T_{[a_1\cdots a_p]}(J_{[a_1\cdots a_p]}+\bar{J}_{[a_1\cdots a_p]})
    \,,
\end{gather}
since $T$ is fully antisymmetric and thereby extracts the fully antisymmetric part of $J$ and $\bar{J}$. Further, because all exchanges of $\psi$ and $\psib$ are compensated by the antisymmetric exchange of the indices
\begin{equation}
    \psi_{[a_1}\cdots \psi_{a_{p-1}}\psib_{a_p]}=\frac{1}{p}\left(\psi_{a_1}\cdots \psi_{a_{p-1}}\psib_{a_p}+\psi_{a_1}\cdots\psi_{a_{p-2}} \psib_{a_{p-1}}\psi_{a_p}+\cdots+\psib_{a_1}\psi_{a_{2}}\cdots\psi_{a_p}\right)
\end{equation}
and the other terms with an interchanged position of $\psib$ bring the same contribution, so in total
\begin{gather}
    J_{[a_1\cdots a_p]}=
    \psi_{a_1}\cdots \psi_{a_{p-1}}\psib_{a_p}+\psi_{a_1}\cdots\psi_{a_{p-2}} \psib_{a_{p-1}}\psi_{a_p}+\cdots+\psib_{a_1}\psi_{a_{2}}\cdots\psi_{a_p}
    =J_{a_1\cdots a_p}
\end{gather}
and similarly for $\bar{J}$.
After the average over $T$, we obtain
\begin{align}
    J_{[a_1\cdots a_p]}\bar{J}_{[a_1\cdots a_p]}=&\left(\psi_{a_1}\cdots \psi_{a_{p-1}}\psib_{a_p}+\psi_{a_1}\cdots\psi_{a_{p-2}} \psib_{a_{p-1}}\psi_{a_p}+\cdots+\psib_{a_1}\psi_{a_{2}}\cdots\psi_{a_p}\right)\notag\\
    &\times\left(\psib_{a_1}\cdots \psib_{a_{p-1}}\psi_{a_p}+\psib_{a_1}\cdots\psib_{a_{p-2}} \psi_{a_{p-1}}\psib_{a_p}+\cdots+\psi_{a_1}\psib_{a_{2}}\cdots\psib_{a_p}\right)\\
    =&(\psib_{a_1}\psi_{a_1})\cdots(\psib_{a_p}\psi_{a_p})\left((-1)^{\frac{p(p+1)}{2}-1}p\right)\,.
\end{align}
Including as well the $\bar{J}J$ term, we are left with the effective action \eqref{eq:effaction2}.

\end{itemize}

\newpage 

\let\oldbibliography\thebibliography 
\renewcommand{\thebibliography}[1]{\oldbibliography{#1}
\setlength{\itemsep}{-1pt}}
\bibliographystyle{JHEP}
\bibliography{biblio.bib}

\end{document}